\documentclass[12pt]{article}
\usepackage{epsfig,psfrag,citesort}
\def \ov {\over }

\newcommand{\cN}{{\cal N}}

\def\ba{\begin{eqnarray}}
\def\ea{\end{eqnarray}}
\def\nn{\nonumber}
\newcommand{\be}{\begin{equation}}
\newcommand{\ee}{\end{equation}}
\newcommand{\bea}{\begin{eqnarray}}
\newcommand{\eea}{\end{eqnarray}}

\makeatletter
\@addtoreset{equation}{section}
\makeatother

\def\appendix#1{
  \setcounter{section}{1}
  \setcounter{equation}{0}
  \renewcommand{\thesection}{\Alph{section}}
  \section*{Appendix \thesection\protect\indent \parbox[t]{11.715cm} {} }
  \addcontentsline{toc}{section}{Appendix \thesection\ \ \ }}

\def\cO{{\cal O}}
\def\tO{{\tilde O}}

\def\bfg{{\,{{{\gamma}\!\!\!\!\hskip.6pt{\gamma}}\!\!\!\!\hskip1.3pt\gamma}\,}}

\font\mybb=msbm10 at 12pt

\def\bb#1{\hbox{\mybb#1}}

\def\T {\bb{T}}
\def\S {\bb{S}}
\def\C{\bb{C}}
\def\PP {\bb{P}}
\def\id{\protect{{1 \kern-.28em {\rm l}}}}

\begin{document}

\begin{titlepage}
\begin{flushright}
NSF-ITP-02-92\\
ITEP-TH-52/02
\end{flushright}

\begin{center}
{\Large $ $ \\ $ $ \\
A Calculation of the plane wave string Hamiltonian  from 
${\cal N}=4$ super-Yang-Mills theory
}\\
\bigskip
\bigskip
David~J.~Gross$\;^{\flat}$, 
Andrei~Mikhailov$\;^{\flat}$\footnote{On leave from 
the Institute of Theoretical and 
Experimental Physics, 117259, Bol. Cheremushkinskaya,
25, 
Moscow, Russia.}, Radu~Roiban$\;^{\sharp}$\\
\bigskip\bigskip\smallskip
$^{\flat}\;$ Kavli Institute for Theoretical Physics,\\
University of California, Santa Barbara, CA 93106\\
\bigskip
$^{\sharp}\;$ Physics Department,\\
University of California, Santa Barbara, CA 93106\\
\vskip .8cm
E-mail: $^{\flat}\;$gross@kitp.ucsb.edu,
andrei@kitp.ucsb.edu,
 $\;\;^{\sharp}\;$radu@vulcan.physics.ucsb.edu
\end{center}
\vskip .8cm

\abstract{
 Berenstein, Maldacena, and Nastase have proposed, as a limit of the
strong form of the AdS/CFT correspondence, that string theory in a
particular plane wave background is dual to a certain subset of operators
in the ${\cal N}=4$ super-Yang-Mills theory. Even though this is {\it a
priori} a strong/weak coupling duality, the matrix elements of the string
theory Hamiltonian, when expressed in
gauge theory variables, are analytic in the 't Hooft coupling constant.
This allows one to conjecture that, like the masses of excited string
states, these can be recovered using perturbation theory  in Yang-Mills
theory.

In this paper we identify the difference between the generator of scale
transformations and a particular $U(1)$ R-symmetry generator as the
operator dual to the string theory Hamiltonian for nonvanishing
string coupling. We compute its matrix elements
and find that they agree with the string theory prediction provided that the
state-operator map is modified for nonvanishing string coupling. We
construct this map explicitly  and calculate  the anomalous dimensions of
the new operators. 
We identify the component arising from the modification of the
state-operator 
map with the contribution of the string theory contact terms to the masses 
of string states.}

\end{titlepage}

\section{Introduction}

{\it ``How do strings behave for strong (world sheet) coupling?''} This question is 
an interesting one and its 
answer could lead to enormous progress in string theory. The various dualities
of string theory were important sources of insight. Perhaps the closest we have come to answering 
this question in 
a particular context comes from the AdS/CFT duality which seems to suggest that string theory on an 
$AdS$ space at large world sheet coupling has a description in form of perturbative $\cN=4$ 
Yang-Mills theory. Proving this statement turns out to be unexpectedly hard due 
to the nonlinearities of the world sheet action. 

It is clear that the above setup is just a limit of the strong form of the AdS/CFT duality. 
Fortunately, other limits can be found in which the action becomes quadratic in light-cone gauge.
An interesting proposal was put forward in \cite{BMN}, relating a particular sector of an $SU(N)$ 
gauge theory to string theory in a plane wave background. The plane wave considered in 
\cite{BMN} 
\be
ds^2=-dx^+dx^--\mu^2(x^2+y^2)dx^+{}^2+dx^2+dy^2~~~~F_{+1234}=F_{+5678}=\mu
\label{wave}
\ee
was obtained as a Penrose limit of $AdS_5\times S^5$. On the gauge theory side the limit focuses 
on the set of operators with 
$R$-charge $J$ which, for fixed Yang-Mills coupling, scales with $N$ such that $J/\sqrt{N}$, as
well as the difference between the dimension of operators and their $R$ charge are fixed.
This setup came to be known as the BMN limit.
The details of this limit imply that, while this proposal is a limit of the strong form of the  AdS/CFT 
duality, it cannot be reached by a sequence of theories each at infinite $N$. It is therefore clear that 
this duality probes finite $N$ regions of the initial AdS/CFT duality. 

Even though it is sourced by a nontrivial RR field, the background (\ref{wave}) is substantially
simpler that the original $AdS_5\times S^5$. It turns out that, for a string in such a background, 
the world sheet theory is free and thus there is no difficulty in quantizing it and finding the free
spectrum \cite{METS},\cite{METY}. The details of the BMN limit relate various operators on the 
string theory  side with operators on the gauge theory side. Of particular relevance for this paper 
is the Hamiltonian
$P^-$ which, for the Euclidean gauge theory, turns out to be given by the difference between the 
generator of scale transformations and the R-charge generator $J$ \cite{BMN}:
\be
P^-=\Delta-J~~.
\label{BMNhamilt}
\ee
From the considerations in \cite{BMN}, it follows that this relation holds at tree level. 

At first sight this limit does not seem to bring us any closer to understanding how a string theory
can be related to a weakly coupled field theory, since the Yang-Mills coupling $g_{YM}$ is
kept fixed while the number of colors is infinite and thus the 't Hooft coupling 
$\lambda=g_{YM}^2N$ is infinite. However, a closer inspection of string theory results reveals that 
they have a weak Yang-Mills coupling expansion as well. This expansion also implies that 
the effective coupling is not the 't Hooft coupling, but $\lambda'=\lambda/J^2$. In string theory 
variables this is $\lambda'=1/(\mu p^+ \alpha')^2$. In the BMN
limit this quantity is arbitrary and can be taken to be small. It appears therefore that in the BMN
limit the gauge theory develops an effective coupling in tems of which it is weakly coupled.
In \cite{GMR} we checked that a finite $N$ and $J$ version of $\lambda'$ is indeed the effective 
field theory coupling to two loops and conjectured that this is the case to all loop orders. Based 
on this conjecture we proved that the anomalous dimensions of operators conjectured to be dual 
to string states have the precise value
predicted by string theory. Our conjecture was later proven in \cite{SAZA} where a different 
computation of the anomalous dimensions was presented. 
This beautiful proof, based on superconformal invariance and using
equations of motion,
represents a strong test of the validity of equations of motion at the
quantum level and from certain points of view suggests that  the matching
of non-BPS operators in the particular subsector of the $\cN=4$ SYM  might be
a consequence of the superconformal symmetry.

Recovering the exact tree level string spectrum seems like a small miracle, since there does not 
exist any {\it a priori} reason for the gauge theory perturbation theory to give the full answer.
It is not clear why the series representation of the anomalous dimension of the operators considered
should converge at all. Nevertheless, the results of perturbative computations in gauge theory seem 
to suggest that string theory in the  plane wave background has a good 
chance of having a description in the form of a field theory which is effectively weakly coupled.

Given the success in matching gauge theory operators with string states, it is natural to wonder
whether string interactions have a perturbative gauge theory representation. The simplicity of 
(\ref{wave}) allows one to explicitly study interactions, splitting and joining of strings. In 
\cite{SPVO} the 3-string hamiltonian was constructed, together with the order $g_s$ 
corrections to 
all dynamical symmetry generators. 

However, the pure $\lambda'$ expansion in gauge theory does not carry any information about 
string interactions. This is clear, since in string theory variables $\lambda'$ is independent on the
string coupling constant. On the string theory side there exists  a dimensionless, arbitrary 
parameter, which is related to the string coupling:
\be
g_2=g_s(\mu p^+ \alpha')^2
\ee
which in gauge theory variables is
\be
g_2={J^2\ov N}~~.
\ee
The appearence of a $1/N$ implies that this parameter describes the genus expansion in gauge 
theory. This is indeed the case, as it has been shown in \cite{MITH}, \cite{SEST}.

Based on gauge theory computations, the authors of \cite{MITH} were led to conjecture a certain
expression for the matrix element of the string Hamiltonian between a 1- and a 2-string state.
The essential test for this conjecture was the recovery of the gauge theory prediction for the
masses of string states at order $g_s$ from a ``unitarity computation'':
\be
\Delta E_n = \sum_{n\ne m}{|V_{nm}|^2\ov E_n- E_m}
\ee
On the string theory side this corresponds to a 1-loop computation.

Subsequent string field theory analysis performed in  \cite{SPVOIIv3} 
revealed that this conjecture does not hold, and the correct matrix 
elements of the string Hamiltonian were computed\footnote{The original version of 
this paper was refering to the second version of hep-th/0206073 where the conjecture 
was confirmed. This does not imply that the original results were technically incorrect, but only that the operator basis constructed in the original version of this paper is not appropriate for comparison with the correct string theory predictions.}.

%
%

The issue of recovering the interactions of the string in a plane wave background from perturbative 
Yang-Mills theory was discussed by various authors: n-point functions were discussed in 
\cite{CHU0}, \cite{CHUI}, \cite{KOREANS} where some tests of the conjectured matrix elements of
string Hamiltonian  were proposed. Computations of n-point functions via unitarity sums was 
discussed in \cite{Huang}.  Some properties of vector operators were analyzed in  \cite{GU} where
an apparent mismatch with unitarity computations was pointed out. A possible solution was suggested
in \cite{CHU}. Deformations of the world sheet action by exactly marginal operators
were discussed in \cite{PARY}, where $1/R_{AdS}$ corrections to the masses of the string states were
analyzed. 

In this paper we will recover the expression for the matrix elements of the string Hamiltonian
from a perturbative gauge theory computation. In particular, one of the main results will be that
equation (\ref{BMNhamilt}) holds to second order in the string coupling constant. In gauge theory 
these matrix elements arise as two-point functions between appropriately-defined (multi-trace) operators.
This result seems to support the suggestion in \cite{HVER} that string interactions are encoded 
in two-point functions of appropriately-defined multi-trace operators. An essential ingredient 
in our derivation are the $g_s$ corrections to the state-operator map of \cite{BMN}. These corrections are 
partially fixed by the requirement that in the new basis the scalar product of operators is the identity matrix.
However, while some redefinitions may seem more natural than others, the requirement that the scalar 
product be the identity matrix fixes the state-operator map up to an $SO(2J+1)$ transformation. This freedom
must be fixed by comparing with the string theory predictions.

In the next section we will identify the Hamiltonian and present arguments that the correspondence
between string states and gauge theory operators must be modified for finite string coupling. We
will also discuss the relation between the anomalous dimension matrix, the scalar product of operators
and the matrix elements of the proposed Hamiltonian. Section 3 is devoted to the computation of the 
anomalous dimension matrix for operators corresponding to string states created by exactly two 
creation operators. The results are summarized in equation (\ref{gamma}). 
In section 4, by comparing the matrix elements of the gauge theory version of the 
string Hamiltonian with the string theory predictions, we construct the operators that are dual at order $g_s$
to string states, equation (\ref{newoperators}). The anomalous dimension of these 
operators receives an extra contribution compared to 
the dimension of the original operators, equation (\ref{contact}). We interpret it as arising from 
the contact terms needed at order $g_s^2$ for the closure of the symmetry algebra. As a byproduct 
of this construction we find the free 2-point functions of double-trace operators at genus 1, equation
(\ref{2pfdouble}). As a 
consistency check we find that the masses of the 2-string states do not receive corrections at order 
$g_s^2$. We conclude with a discussion of our results and directions for future work.

{\bf Note:} While this paper was being written we received the
interesting paper \cite{SESTII} discussing mixing of operators. While 
we agree with the reason for this mixing, we depart  from
the particular implementation discussed in that paper. We find that
recovering the matrix elements of the string Hamiltonian from a gauge
theory standpoint requires a different choice for the 
state-operator map, which is related to the one described in \cite{SESTII}
by an orthogonal transformation.

{\bf Note for the second version:} In the original version of this paper we constructed 
the state-operator map such that the matrix elements of the Hamiltonian computed in gauge 
theory agree with the string theory results at the time the paper was written, hep-th/0206073. 
The idea of the construction is of course unchanged, but due to subsequent corrections to the 
string theory computation, this version of the paper presents a different state-operator map. 
The two sets of operators are related by an $SO(2J+1)$ transformations mentioned above as 
well as later in the text. The freedom of performing such transformations was part of the initial 
construction.


\section{Hamiltonian, operators and mixing }

We begin this section with a fast review of the BMN operators and their associated 
string states. To keep the discussion self-contained and fix our conventions and 
notations we list here the bosonic part of the ${\cal N}=4$ Lagrangian with the 
normalizations we will use in our computations.
\be
{\cal L}={1\ov g_{YM}^2}Tr\Bigg[
D_\mu{\Phi}^i D^\mu{\Phi}_i + D_\mu{\bar Z}^i D^\mu{Z}_i + 
[Z,\,\Phi_i][{\bar Z},\,\Phi^i]-{1\ov 4}[Z,\,{\bar Z}]^2+{1\ov 2}[\Phi_i,\,\Phi_j][\Phi^i,\,\Phi^j]
\Bigg]
\label{lagrangian}
\ee
with $i=1,\dots,4$ and $Tr [T^a T^b]={1\ov 2}\delta^{ab}$. The field $Z$ is complex and 
corresponds to singling out a $U(1)$ subgroup of the $SU(4)$ R-symmerty group, $Z=\Phi^5+i\Phi^6$,
while the fields $\Phi_i$ are real.

A generic BMN operator is indexed by a set of $n$ integers $m_i,~i=1,...,n$ and has 
the expression:
\be
{\cal O}^{i_1\dots i_n}_{m_1\dots m_n}={1\ov N^{(J+n+1)/2}f(J,n)}
\sum_{\stackrel{l_0,\dots ,l_n=0}{l_0+\dots+l_n=J}}^J
\left(\prod_{k=1}^ne^{2\pi i (l_0+\dots+l_k){m_k\ov J}}\right)
Tr[\Phi_{0}Z^{l_0}\prod_{j=0}^{n}(\Phi_{i_j}Z^{l_j})]
\label{BMNoperators}
\ee
where $i=1,\dots ,4$ and the function $f(J,n)$ is chosen such that the free 2-point function of these operators 
is ${1\ov |x-y|^{2(J+n+1)}}$.
It turns out that this operator has finite anomalous dimension in the BMN
limit, to all orders in the Yang-Mills coupling \cite{GMR}, \cite{SAZA}:
\be
\gamma_{m_1\dots m_n}=\sum_{i=1}^n\left(\sqrt{1+ 16\pi^2 m_i^2 \lambda'}-1\right)~~.
\ee

These operators (\ref{BMNoperators}) are conjectured to be in one to one correspondence 
to the string states in plane wave background, as follows:
\be
{\cal O}^{i_1\dots i_n}_{m_1\dots m_n}~~~\leftrightarrow~~~~
a^\dagger{}^{i_0}_{-\sum_i m_i} \prod_{j=1}^n a^\dagger{}^{i_j}_{m_j}|0\rangle 
\ee
where $|0\rangle $ is the lightcone vacuum and $a^\dagger {}_m^i$ is the creation operator
for the $i^{\rm th}$ bosonic string coordinate, with frequency $\mu\sqrt{1+ m^2/(\mu p^+\alpha')^2}$. 
This correspondence will be modified at finite string coupling as we will discuss shortly.


\subsection{The Hamiltonian}

As described in the introduction, the operator dual to the tree-level string theory 
Hamiltonian $P^-$ is the difference between the generator of scale transformations
and the generator of $R$-symmetry transformations along the geodesics chosen
for the Penrose limit. This identification is in fact based on the symmetries of the 
plane wave which are just a contraction of the 4-dimensional superconformal group.
This, together with the fact that the plane wave is an exact solution of string theory, 
implies that the identification between the string Hamiltonian and the difference between
the generator of scale transformations and a $U(1)$ $R$-symmetry generator survives
as we turn on the interactions. Thus, we expect the following relation to hold:
\be
P^-(g_s) =\Delta-J~~.
\label{hamilt}
\ee
The purpose of this paper it to show that this equation holds to leading order in the 
string coupling. 

To prove that two operators are the same it is enough to show that they have the 
same eigenvalues. This guarantees that, for any choice of basis for the
carrier space of one of them, there exists a choice of basis for the carrier space of the other one 
such that their matrix elements agree. Thus, from this perspective,
proving that two operators are the same involves constructing a map between the spaces
they act upon as well as showing that their matrix elements in the basis related by this map,
are the same. Thus, to reach our goal it is necessary to construct a map between 
the string and gauge theory Hilbert spaces, i.e. we have to find the operators 
dual to multi-string states as well as the appropriate scalar products. We stress that it is not possible
to construct this map without performing a detailed comparison of the matrix elements of some 
operator. As described in the 
previous subsection, it was argued in \cite{BMN} that, at vanishing string coupling, the operators 
dual to 1-string states are the BMN operators (\ref{BMNoperators}).
Explicit gauge theory computations showed that indeed, at vanishing string coupling, 
the anomalous dimensions of these operators match the tree-level string state masses.

The operators dual to 2-string states can be infered rather easily. In string theory a 2-string 
state belongs to the tensor product of two 1-string Hilbert spaces. It is therefore natural to 
identify, at vanishing string coupling, the 2-string states as being the double-trace operators 
obtained by simply multiplying two BMN operators. 

The next step in constructing the map is the identification of a scalar product. On the 
string theory side the scalar product is fixed to be the scalar product of the 1-string Hilbert 
space and tensor products thereof for multistring Hilbert spaces. On the gauge theory side,
a natural candidate for the result of the  scalar product of two operators is the coefficient of 
their 2-point function. 
There is, however, a substantial difference between this and the string theory scalar product.
While the latter is diagonal to all orders in the string coupling, the former receives various
corrections. Thus, the identification between the string and gauge theory scalar products
must be reanalyzed order by order $g_2$ expansion in the gauge theory. This implies that, unless all 2-point 
functions are trivial, the state-operator map is bound to be modified at finite order 
in $g_s$. For example, the free 3-point function of BMN operators of certain types was computed
in \cite{SEST}\cite{MITH}  and found to be nontrivial at order $g_2\sim g_s$. 
It follows that the gauge theory scalar product between operators dual to 1- and 2-string 
states does not vanish at order $g_2\sim g_s$. This fact is crucial for the correct identification
of the matrix elements of the Hamiltonian. We will therefore compute the matrix elements of
equation (\ref{hamilt})
\be
\langle {\bar {\cal O}}_i[(\Delta-J){\cal O}_{jk}]\rangle=
\langle i|P^- |j\rangle|k\rangle\equiv (\langle i|\otimes \langle j|\otimes \langle k|)|V\rangle
\label{mith}
\ee
where ${\cal O}_i$ and ${\cal O}_{jk}$ are the operators\footnote{This notation will be shortly
changed to a (hopefully) more convenient one.} dual to the 1- and 2-string states
$|i\rangle$ and $|j\rangle|k\rangle$ respectively.

\subsection{The anomalous dimension matrix}

Using the generator of scale transformations as the Hamiltonian produces an intimate 
relation between the anomalous dimension matrix and the matrix elements of the 
Hamiltonian. Thus, let us proceed with a brief  summary of the properties of
anomalous 
dimension matrix.

Consider a set of operators ${\cal O}_\Sigma$ \footnote{Of particular interest for us is the case when 
this is the set of all single- and double-trace operators with $\Delta-J=2$ at tree level. This distinction 
is irrelevant at this point.}.  An aspect of the renormalization of 
composite operators 
is operator mixing: the divergences in 1PI diagrams with one insertion 
of a composite operator generally contain divergences proportional to 
other composite operators. Thus, all composite operators must be
renormalized at the same time, order by order in perturbation theory.
This leads to the following relation:
\begin{equation}
{\cal O}_\Sigma^{bare}(\Phi^{bare},g^{bare})=\sum_B\,Z_\Sigma{}^\Xi
{\cal O}_\Xi^{ren}(Z_3^\Phi\Phi^{ren}, {Z_1\over Z^\Phi_3{}^{3/2}}g^{ren})
\label{renormalization}
\end{equation}
where $\Phi$ denotes a generic field, we assumed the existence of 
a cubic $\Phi$ interaction of strength $g$ and $Z_3$ and $Z_1$ are the usual
wave function and coupling constant $Z$-factors.

This implies that under scale transformations, the operators $\cO_\Sigma$ 
transform as
\be
\cO_\Sigma~~\longrightarrow~~\Delta {\cal O}_\Sigma=\sum_\Xi(d_\Sigma\delta_{\Sigma\Xi}+
\gamma_{{}_{\Sigma\Xi}}){\cal O}_\Xi
\label{scale}
\ee
were $\bfg$ is the anomalous dimension matrix, and $d_\Sigma$ is the engineering dimension of 
$\cO_\Sigma$. Its expression, 
in terms of the renormalization factors $Z_\Sigma{}^\Xi$ introduced above, using dimensional regularization,
is given by:
\be
\gamma_{{}_{\Sigma\Xi}}(\lambda)=\lim\limits_{\epsilon\rightarrow 0}\left[
\epsilon \,(Z^{-1}(g_{YM},\,\epsilon))_\Sigma{}^\Psi
{d\ov d\ln g_{YM}}
Z_{\Psi}{}^{\Xi}(g_{YM},\,\epsilon)\right]
\ee
where $g_{YM}$ is the Yang-Mills coupling. 
We will work mostly at 1-loop level, in which case the expressions simplify 
considerably. The renormalization constant $Z_Sigma{}^\Xi$ is 
\be
Z_{\Sigma\Xi}=\delta_{\Sigma\Xi}+{g_{YM}^2\ov \epsilon}Z^{(1)}_{\Sigma\Xi}
\ee
and thus the anomalous dimension matrix is
\be
\gamma_{{}_{\Sigma\Xi}}=2g_{YM}^2\,Z^{(1)}_{\Sigma\Xi}~~.
\label{anomdimdef}
\ee
Thus, at this level, the entries of the anomalous dimension matrix can be determined 
in terms of the coefficient of the $1/\epsilon$ in 1-loop diagrams with one insertion of 
the composite operator.

As pointed out before, the relation between the matrix elements of the Hamiltonian 
and the anomalous dimension matrix depends on the scalar product. More explicitly, 
assume that 
\be
\langle {\bar{\cal O}}_\Sigma(x) {\cal O}_\Xi(y)\rangle=S_{\Sigma\Xi}{1\ov  |x-y|^{d_\Sigma+d_\Xi}}~~,
\ee
or schematically, as a scalar product,
\be
\langle {\cal O}_\Sigma {\cal O}_\Xi \rangle=S_{\Sigma\Xi}~~.
\ee
Then, the matrix elements of $\Delta-J$ are given by
\be
\langle {\cal O}_\Sigma(\Delta-J){\cal O}_\Xi\rangle=\sum\Psi(\gamma_{{}_{\Xi\Psi}}+
(d_\Xi-J_\Xi)\delta_{\Xi\Psi})S_{\Sigma\Psi}
\label{matelem}
\ee
The identification of $\Delta-J$ with the string Hamiltonian $P^-$, requires that it be Hermitian.
This is reasonable since $\Delta$ is the ``Hamiltonian'' of the Euclidian theory using radial
quantization. On the other hand, $\bfg$ needs not be Hermitian. The above equation
resolves this concern  about the interpretation of $\Delta$, requiring only that $S\bfg^T$ be Hermitian.

 As we will see in the following sections, for the operators 
we are interested in, the anomalous dimension matrix is {\it not} symmetric but the 
Hermiticity of the Hamiltonian is recovered.


\subsection{Divergences\label{secdivergences}}

Before we proceed with the details of the computation for operators dual to massive 
states  generated by two creation operator let us describe the various ingredients that
enter, not only in this computation, but also in the ones required for general operators.

The BMN limit is a large $N$ limit which is different from the usual 't Hooft limit. In the
't Hooft limit perturbation theory is organized as a double series with expansion parameters
$\lambda=g_{YM}^2N$ and $1/N$. The first one counts the loop order of the diagram while 
the latter counts the genus of the diagram. For large $N$ high powers of $1/N$ can be safely 
neglected, and the dominant contribution to all processes comes from planar diagrams.
Similarly, in the BMN limit there are two expansion parameters: 
$\lambda'=g_{YM}^2N/J^2$  and $g_2=J^2/N$. As in the 't Hooft limit, the exponent of
$\lambda'$  counts the number of loops in the diagram. However, now $g_2$ counts the genus.
In the limit in which the string theory is supposed to be described as a gauge theory this parameter
 is of order one, and thus higher genus diagrams cannot be neglected. 

On the string theory side the perturbation theory is organized following the genus of the Feynman 
diagrams, with coefficients $g_s^{\chi}$, where $\chi=2g-2$ is the Euler number of the diagram.
Since $g_2$ expressed in string theory variables is proportional to $g_s$, it follows that higher
genus gauge theory diagrams contain informations about string perturbation theory.

From this discussion it is clear why the mixing of single- and double-trace operators is 
parametrically relevant for the string splitting and joining. First, the mixing in question 
starts occurring at 1-loop; thus, it will be proportional to $\lambda'$.  Then, diagrammatically,
the relevant Feynman diagrams can be drawn on a sphere with three holes, a pair of pants.
This counts as  ``genus $1/2$'' and thus the amplitude will be proportional to $g_2$. Expressing this 
product in string theory variables we find that $\lambda'g_2=g_s$, 
which is the correct weight of the 3-string vertex.

\begin{figure}[h]
\begin{center}
\epsfig{file=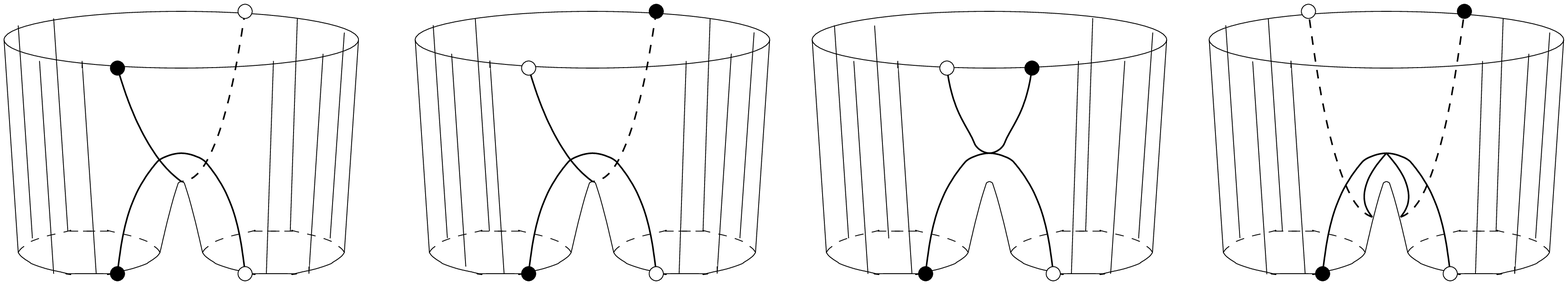,width=15cm}
\caption{Feynman diagrams leading to the mixing of double-trace operators with single-trace operators,  
drawn on a pair of pants.\label{algmFdpants}}
\end{center}
\end{figure}

We now describe in more detail the Feynman diagrams that will lead to the entries of the
anomalous dimension matrix describing the mixing of  single- and double-trace operators. 
Let us consider a double-trace operator and let us draw each of the traces composing it as
a circle. Since we want to find the admixture of single-trace operators in the scale transformations
of this operator, we are interested in the Feynman diagrams, with one insertion of  a single trace operato, so that   
all external lines  appear in  the same trace. At 1-loop level only two fields 
can participate in the 
interaction. If both of them belong to the same trace, then such diagram will produce the anomalous dimension of the double-trace operator, and the mixing with single-trace ones 
is only due to the tree-level mixing of these operators. It is therefore clear that, to find the mixing due to 
interactions, the two interacting fields must belong to different traces of the original operator. All relevant Feynman 
diagrams with this property are drawn, on a sphere  with three holes, in Figure \ref{algmFdpants}. 
In these diagrams the fields taking part in the interactions are denoted by an empty and a filled circle; they 
can be either $Z$-fields or impurity fields.
The spectator fields  are denoted by straight lines and the top circle denotes the fact that outgoing fields
are all under the same trace. These are all 1-loop diagrams since all fields in the double-trace operator 
are  at the same space-time point. There are more 1-loop diagrams that take a double-trace operator 
and transform it in a single-trace one. Examples of such diagrams can be obtained by interchanging 
the black and the white dots on the top circles in the diagrams in  figure \ref{algmFdpants}. However, as 
we will see shortly, only the diagrams in figure \ref{algmFdpants} are divergent and thus they are the 
only ones that contribute to the anomalous dimension matrix. 

\begin{figure}[ht]
\begin{center}
\epsfig{file=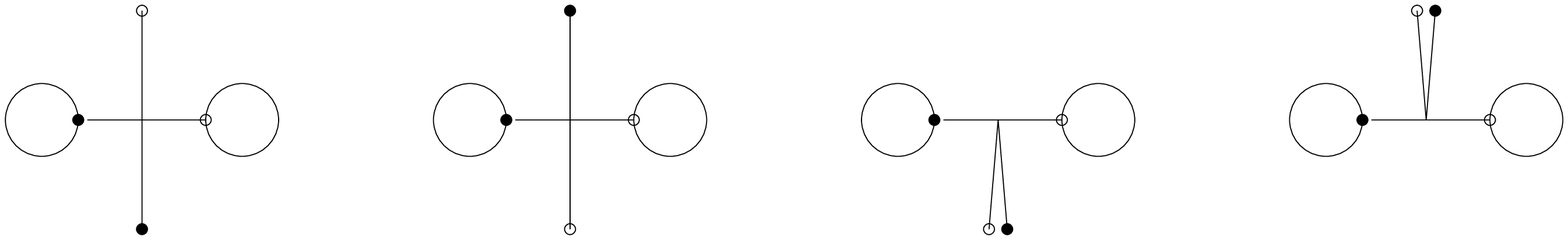,width=15cm}
\caption{Symbolic representation of the diagrams in figure \ref{algmFdpants}.\label{algmFd}}
\end{center}
\end{figure}

\begin{figure}[ht]
\begin{center}
\psfrag{Z}{$\scriptstyle{Z}$}
\psfrag{A}{$\scriptstyle{\Psi}$}
\psfrag{B}{${\scriptstyle -2\times}{1\ov\epsilon}$}
\psfrag{C}{${1\ov\epsilon}$}
\psfrag{D}{${1\ov\epsilon}$}
\psfrag{I}{${\scriptstyle -}{1\ov\epsilon}$}
\psfrag{K}{${\scriptstyle -}{1\ov\epsilon}$}
\psfrag{G}{${1\ov\epsilon}$}
\psfrag{H}{${1\ov\epsilon}$}
\psfrag{L}{${\scriptstyle -}{1\ov\epsilon}$}
\psfrag{M}{${\scriptstyle -}{1\ov\epsilon}$}
\psfrag{N}{${\scriptstyle 2\times}{1\ov 2 \epsilon}$}
\psfrag{O}{${\scriptstyle 2\times}{1\ov 2 \epsilon}$}
\psfrag{ZZ}{$\scriptstyle{\Phi}$}
\epsfig{file=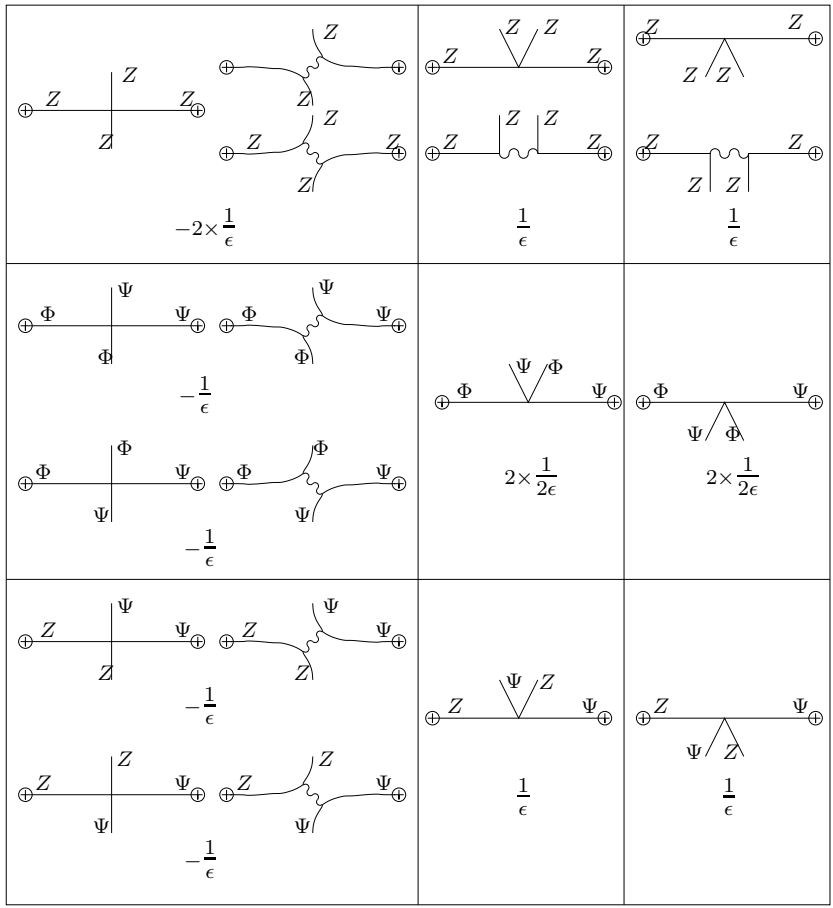,height=12cm}
\caption{Divergent 1-loop diagrams.\label{divgraphs}}
\end{center}
\end{figure}

Let us now introduce a convenient, schematic notation for these diagrams, which reduces the problem of
computing them to just a combinatorial one. As above, we denote a trace  by one circle 
and the fields that interact by empty and filled dots. We will omit the spectator fields as well as the fact that
the outgoing fields are under the same trace. Then, we represent the  four diagrams above as in figure \ref{algmFd}.
They are just projections of the diagrams in figure \ref{algmFdpants} on the plane containing the 
two circles denoting the double-trace operator. To turn this in a combinatorial problem we have to 
compute all the divergent 1-loop diagrams and then identify the black and white dots. The results of this
computation are summarized in figure \ref{divgraphs}.

The results stated in figure \ref{divgraphs} include their combinatorial factors 
arising from the Lagrangian. The points denoted by $\oplus$ can be either a black dot or a white dot.
Thus, computing all diagrams represented in figure \ref{algmFdpants} reduces to finding all possible ways 
of linking the two traces building a double-trace operator using links appearing in figure \ref{divgraphs}.

\begin{figure}[ht]
\begin{center}
\epsfig{file=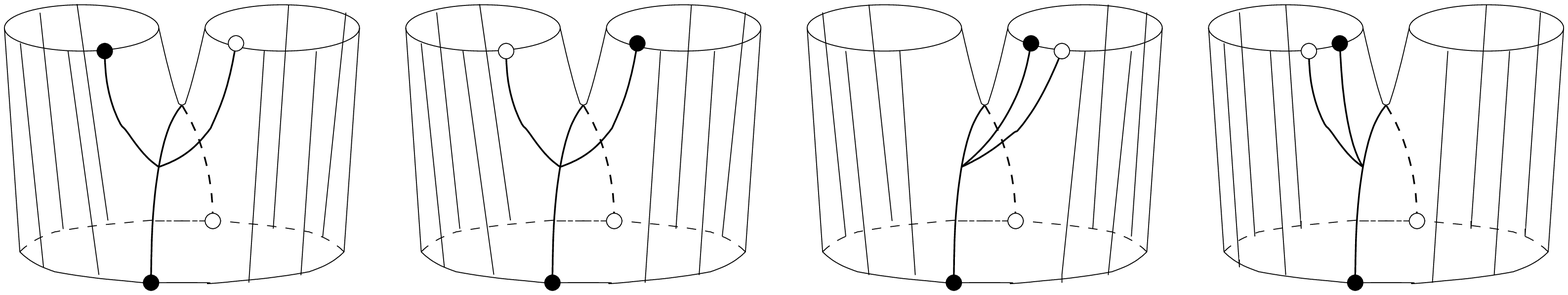,width=15cm}
\caption{Feynman diagrams leading to the mixing of single-trace operators with double-trace 
operators, drawn on a pair of pants.\label{algcoFdpants}}
\end{center}
\end{figure}

A similar discussion holds for the diagrams leading to the mixing of single-trace with 
double-trace operators. We start with a single-trace operator, denoted by one circle, while the fields 
participating in the interaction are denoted by a black and a white dot. Turning
it into a double-trace operator via an interaction vertex requires that the participating fields
are not neighbors in the original operator. If they were, then the Feynman diagrams would yield
the anomalous dimension of the single-trace operator and the mixing with the double-trace operators
arises because of the tree-level mixing. For our purpose, the relevant Feynman diagrams, drawn on 
spheres with three holes, are listed in figure  \ref{algcoFdpants}. The spectator fields are again represented 
by straight lines.

\begin{figure}[ht]
\begin{center}
\epsfig{file=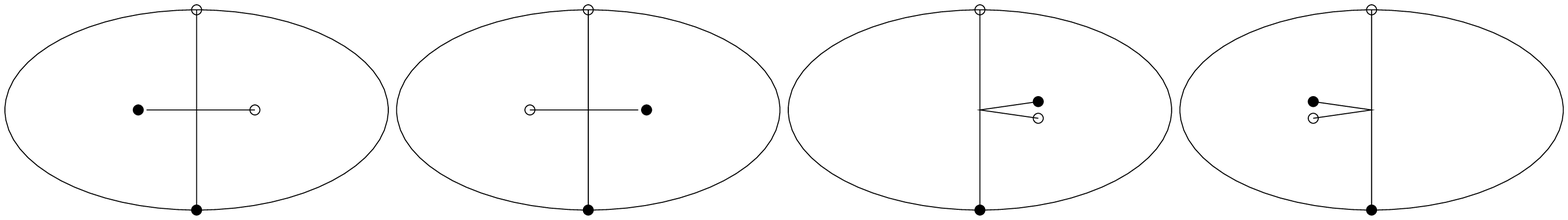,width=15cm}
\caption{Symbolic representation of the diagrams in figure \ref{algcoFdpants} .\label{algcoFd}}
\end{center}
\end{figure}

These diagrams can be also reduced to a combinatorial problem, as in the case of the mixing of 
double- with single-trace operators.  With the same notation for the trace and for the interacting fields, 
the Feynman diagrams in figure \ref{algcoFdpants} are shown in figure \ref{algcoFd}, respectively.
As before, the spectator fields are not represented. Also, the resulting double-trace operator is 
not drawn explicitly, but it is fairly obvious from the figure. This representation is just a (simplified) projection 
on the plane containing the circle denoting the single-trace operator. As before, the problem of computing 
the Feynman diagrams yielding a double-trace operator out of a single-trace one is reduced to finding
all ways of linking two non-neighboring fields in the single-trace operator using links from figure
\ref{divgraphs}. 

It is worth noting that not all the diagrams in figure \ref{algcoFdpants} are not just the upside-down version
of the diagrams in figure \ref{algmFdpants}. In particular, we notice that inverting the last two diagrams
in figure \ref{algmFdpants} leads to diagrams in which the mixing between single- and double-trace operators
is due to their tree-level mixing and thus it must not be considered in the computation of the anomalous dimension 
matrix.

\begin{figure}[ht]
\begin{center}
\epsfig{file=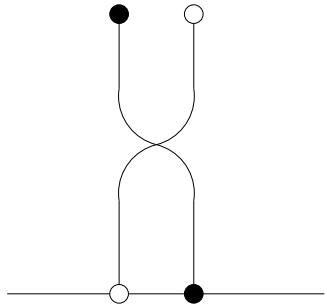,height=4cm}
\caption{Feynman diagrams leading to the anomalous dimension of single-trace operators.\label{example}}
\end{center}
\end{figure}

Before we continue with the main computation, let us demonstrate the use of the building blocks 
in figure \ref{divgraphs}, by rederiving the well-known \cite{BMN}, \cite{GMR}, \cite{MITH}, 
1-loop anomalous dimension of the operator with a single impurity:
\be
\cO=\sum_{l=0}^J e^{il\varphi_m^J}\,Z^l\Phi Z^{J-l}~~.
\ee
In the schematic notation of figure \ref{algmFd} and \ref{algcoFd}, the relevant diagrams are drawn in figure
\ref{example}.
The white and black dots can be either a $Z$ field or the $\Phi$ field. The remaining diagrams, i.e. those that contribute to the 
wave function renormalization of the external lines, are not represented
as they are not needed later on. We will use their expression from \cite{GMR}. 

Using the second entry on the third line  in figure \ref{divgraphs} we find that part of the needed 
counterterm (up to terms arising for finite $J$) is:
\be
\delta_1 \cO= -{1\ov\epsilon}\left[e^{i\varphi_m^J}\cO+e^{-i\varphi_m^J}\cO\right]~~.
\ee
The first term arises by identifying the white dot with a $Z$-field, while the second one from identifying the 
white dot with $\Phi$.

The second part of the counterterm arises by identifying both the black and the white dots with $Z$-fields
and using the second entry on the first line in figure \ref{divgraphs}:
\be
\delta_2 \cO= -{1\ov\epsilon}(J-1)\cO
\ee

The third and last part comes from the wave function renormalization of the external lines. From 
\cite{GMR} we have:
\be
\delta_3 \cO= {1\ov\epsilon}(J+1)\cO
\ee

Combining everything, and restoring the coupling constant dependence, we find the following counterterm:
\be
(\delta_1+\delta_2+\delta_3)\cO={\lambda\ov\epsilon}(2-e^{i\varphi_m^J}-e^{-i\varphi_m^J})\cO
\ee
which leads to the correct anomalous dimension for this operator:
\be
\gamma_{n}=-2\lambda (e^{i\phi}+e^{-i\phi}-2)\simeq 4\lambda\phi^2=
8\pi^2n^2{\lambda\ov J^2}\, .
\label{anomdim}
\ee
To all loop orders and for an arbitrary number of impurities
we found \cite{GMR},\cite{SAZA}, up to boundary contributions:
\be
\gamma_{nn}=\sum_m N_m \gamma_m~~~~~~{\rm where}~~~~~~~~~
\gamma_m=\sqrt{1+16\pi^2 m^2 {g_{YM}^2N\ov J^2}}-1~~~~.
\label{fullanomdim}
\ee

The preliminaries presented in this section can be used to compute the entries of the
anomalous dimension matrix associated to any single- and multiple-trace operators.
In the following we will concentrate on operators which are simple enough for a 
reasonably-sized exposition yet complex enough to exhibit most (hopefully all) possible 
subtleties.


\section{Massive states generated by two creation operator; detailed analysis}

In this section we discuss in detail the computation of the 1-loop anomalous dimension matrix 
for BMN operators with two impurities, i.e. operators that are dual to the simplest massive string 
states created by two creation operators acting on the vacuum. The relevant operators are:
\bea
\cO^{[\Phi,\Psi]_J}_m&=&K^{[\Phi,\Psi]_J}_m
\sum_{l=0}^J e^{2\pi i l \varphi_m^J}Tr[\Phi Z^l\Psi Z^{J-l}]~~~~~\varphi_m^J=
{m\ov J} ~~~~m=0,\dots,J-1\nn\\
{\cal O}^{[\Phi]_{J_1}[\Psi]_{J-J_1}}_{0,0}&=&K^{[\Phi]_{J_1}[\Psi]_{J-J_1}}_{0,0}
Tr[\Phi Z^{J_1}]Tr[\Psi Z^{J-J_1}] \nn\\
{\cal O}^{[\Phi,\Psi]_{J_1}[\id]_{J-J_1}}_{m,\emptyset}&=&
K^{[\Phi,\Psi]_{J_1}[\id]_{J-J_1}}_{m,\emptyset}
\sum_{l=0}^{J_1} e^{2\pi i l \varphi_m^{_1}}Tr[\Phi Z^l\Psi Z^{J_1-l}] Tr[Z^{J_2}] 
\label{double}
\eea
with the normalization constants
$K$ above are given by:
\bea
K^{[\Phi,\Psi]_J}_m&=&{1\ov N^{(J+2)/2}\sqrt{J+1}}~~~~~~~~
K^{[\Phi]_{J_1}[\Psi]_{J-J_1}}_{0,0}={1\ov N^{(J+2)/2}}\nn\\
&&\!\!\!\!\!\!
K^{[\Phi,\Psi]_{J_1}[\id]_{J-J_1}}_{m,\emptyset}={1\ov N^{(J+2)/2}\sqrt{J-J_1}\sqrt{J_1+1}}~~.
\eea
In equation (\ref{double}) refer to any of the four real fields $\Phi_i$ and, for simplicity, 
we will take them to be distinct, $\Phi\ne\Psi$. To lowest order in the string coupling constant
these operators correspond to:
\bea
\cO^{[\Phi,\Psi]_J}_m&{}&\rightarrow~~\alpha^\dagger{}_{-m}^\Phi \alpha^\dagger{}_{m}^\Psi |0\rangle
\nn\\
{\cal O}^{[\Phi]_{J_1}[\Psi]_{J-J_1}}_{0,0}&{}&\rightarrow~~
{1\ov\sqrt{2}}\left(
\alpha^\dagger{}_{0}^\Phi |0\rangle \otimes \alpha^\dagger{}_{0}^\Psi |0\rangle
+
\alpha^\dagger{}_{0}^\Psi |0\rangle \otimes \alpha^\dagger{}_{0}^\Phi |0\rangle
\right)
\nn\\
{\cal O}^{[\Phi,\Psi]_{J_1}[\id]_{J-J_1}}_{m,\emptyset}&{}&\rightarrow~~
{1\ov\sqrt{2}}\left(
|0\rangle \otimes \alpha^\dagger{}_{-m}^\Phi \alpha^\dagger{}_{m}^\Psi |0\rangle
+
\alpha^\dagger{}_{-m}^\Phi \alpha^\dagger{}_{m}^\Psi |0\rangle \otimes |0\rangle
\right)
\eea
Let us comment on the range of the index $m$ in the first and last line in equation (\ref{double}).
There are $J+1$ independent operators  of the type $Tr[\Phi Z^l\Psi Z^{J-l}]$ and out of them we can 
build only $J+1$ independent operators:  $\cO^{[\Phi,\Psi]_J}_m$ with $m$ taking values from 
$0$ to $J-1$ and another operator which is orthogonal on $\cO^{[\Phi,\Psi]_J}_m$ for all $m$ 
and will be discarded. Later on, we will take the limit 
$N,J\rightarrow\infty$ and we will need to sum over the index $m$ and it turns out that we 
have the sum from $-\infty$ to $+\infty$. The reason for this is the following. At finite $N$ and $J$ all 
functions  we encounter are periodic
under $m\rightarrow m+J$, a  typical function is  $\sin^2\pi m/J$. However, once we expand for  
large $J$, we have access 
only to the increasing branch of the sine. To recover the missing piece we  should take $m$ to run from 
$-\infty=\left[{J\ov 2}\right]$ to $+\infty=\left[{J\ov 2}\right]$. A similar 
discussion  applies to ${\cal O}^{[\Phi,\Psi]_{J_1}[\id]_{J-J_1}}_{m,\emptyset}$, except that now
$m$ runs from $\left[{J_1\ov 2}\right]$ to $\left[{J_1\ov 2}\right]$.

We will use the following indices for the anomalous diemnsion matrix and the various transformation 
matrices we will encounter later: a lower-case letter from the middle of the alphabet, say $i$, will 
correspond to a single-trace operator and an upper case letter from the beginning of the alphabet, say $A$,
will denote a double-trace operator. The set of double-trace operators is naturally split in two subsets: the first
subset contains only the operator ${\cal O}^{[\Phi]_{J_1}[\Psi]_{J-J_1}}_{0,0}$ which will be denoted by an 
$x$ index, and the operators  ${\cal O}^{[\Phi,\Psi]_{J_1}[\id]_{J-J_1}}_{m,\empty}$ which will be denoted 
by a lower case greek index from the beginning of the alphabet, say $\alpha$.


We will also use the following notation for the building blocks of BMN operators:
\be
\tO^{[\Phi,\Psi]_J}_l=Tr[\Phi Z^l\Psi Z^{J-l}]
\ee
in terms of which they are: 
\be
{\cal O}^{[\Phi,\Psi]_J}_m=K_m^{[\Phi,\Psi]_J}\sum_{l=0}^Je^{2\pi i n {l\ov J}} 
\tO^{[\Phi,\Psi]_J}_l~~.
\label{BMNop}
\ee

The operators $\cO$ are a kind of discrete Fourier transform of the operators $\tO$.
However, the parallel is not exact at finite $J$, since there are $(J+1)$ independent operators $\tO$
and only $J$ operators $\cO$, since ${\cal O}^J_n={\cal O}^J_{n+J}$. 

To find the expression of $\tO_l^J$ in terms of BMN operators ${\cal O}^{[\Phi,\Psi]_J}_m$, we first notice
that in  ${\cal O}^{[\Phi,\Psi]_J}_m$ only the combination $(\tO^J_0+\tO^J_J)$ appears. Thus, for our 
purpose, we have $J$ operators ${\cal O}$ as linear combinations of $J$ operators $\tO$. The 
relations (\ref{BMNop}) can easily be inverted with the result:
\bea
K^{[\Phi,\Psi]_J}\tO^J_l&=&{1\ov J}\sum_{n=0}^{J-1}e^{-2\pi i\,l{n\ov J}}{\cal O}^{[\Phi,\Psi]_J}_n~~~~~
(\forall)~~~l=1\dots (J-1)\nn\\
K^{[\Phi,\Psi]_J}(\tO^J_J+\tO^J_0)&=& {1\ov J} \sum_{n=0}^{J-1}{\cal O}^{[\Phi,\Psi]_J}_n
\label{invFt}
\eea

However, at infinite $J$ the situation is slightly different. In particular, we can take separately
$K^{[\Phi,\Psi]_J}\tO^J_J$ and $K^{[\Phi,\Psi]_J}\tO^J_0$ to be equal to the right-hand-side 
of the second equation (\ref{invFt}). 

\subsection{Order of limits}

The issue to be addressed is then when is the limit $N\rightarrow\infty$ (which implies $J\rightarrow\infty$ 
according to the BMN limit) supposed to be taken? To answer this question let us go back to the single-trace
operators ${\cal O}^{[\Phi,\Psi]_J}_m$, their scalar product and the boundary contributions to their 
anomalous dimensions. Let us first compute the scalar product at finite $N$ and $J$. It is easy to see that
it is:
\be
\langle{\bar \cO}^{[\Phi,\Psi]_J}_m{\cal O}^{[\Phi,\Psi]_J}_n\rangle=\delta_{mn}+{1\ov J}
\label{scprod}
\ee
This non-orthogonality of order $1/J$ implies that ${\cal O}^{[\Phi,\Psi]_J}_n$ are not eigenvectors
of the generator of scale transformations $\Delta$ at any order in $\lambda'$ and $g_2$ if $J$ and $N$ are finite.
This is the case because $\Delta$ is Hermitian and the eigenvalues that are supposed to correspond to
${\cal O}^{[\Phi,\Psi]_J}_n$ are different for different $n$. A short computation reveals that the boundary 
contributions, i.e. terms due to the collision of $\Phi$ and $\Psi$, spoil the desired properties of the set of
operators ${\cal O}^{[\Phi,\Psi]_J}_n$. Indeed, it turns out that
\bea
\Delta{\cal O}^{[\Phi,\Psi]_J}_m&=&
4g_{YM}^2N\left(4\sin^2\pi\varphi_m^J\right)\cO_m^{[\Phi,\Psi]_
J}
\nonumber\\
&-&2{g_{YM}^2N\over N^{(J+2)/2}\sqrt{J+1}}\left(4\sin^2\pi\varphi_m^J\right)
(\tO^{[\Phi,\Psi]_J}_0+\tO^{[\Phi,\Psi]_J}_J)
\nonumber\\
&-&
2{g_{YM}^2N\over N^{(J+2)/2}\sqrt{J+1}}\left(2i\sin 2\pi\varphi_m^J\right)
(\tO^{[\Phi,\Psi]_J}_0-\tO^{[\Phi,\Psi]_J}_J)
\label{boundary}
\eea
The appearance of the last term is troublesome since its coefficient is not finite as $N$ and $J$ are taken to 
infinity. This problem can be solved as follows.  It is easy to match the construction of string states with the 
construction of representations of the symmetry algebra of the plane wave using creation operators acting
on the light-cone vacuum \cite{METS}. This matching suggests that the number of BMN operators and the
result of their transformations under symmetries of the plane wave matches the number of string states.
All operators that are orthogonal to all the BMN operators can then be discarded. This is the case of 
the operator $(\tO^{[\Phi,\Psi]_J}_0-\tO^{[\Phi,\Psi]_J}_J)$ appearing in the last line of (\ref{boundary}).
This orthogonality is the reason this term did not appear in the 2-point function computations
of \cite{MITH} and \cite{SEST}.

Thus, for our purpose, we can write:
\bea
\Delta{\cal O}^{[\Phi,\Psi]_J}_m&=&
4g_{YM}^2N\left(4\sin^2\pi\varphi_m^J\right)\cO_m^{[\Phi,\Psi]_
J}
\nonumber\\
&-&2{g_{YM}^2N\over N^{(J+2)/2}\sqrt{J+1}}\left(4\sin^2\pi\varphi_m^J\right)
(\tO^{[\Phi,\Psi]_J}_0+\tO^{[\Phi,\Psi]_J}_J)
\eea
This implies that, as stated before, ${\cal O}^{[\Phi,\Psi]_J}_m$ is not an eigenvector of the generator of scale 
transformations, the mismatch being of order ${1\ov\sqrt{J}}$ and therefore it does not 
have a definite anomalous dimension for finite $N$ and $J$. At this stage it seems that the natural solution would be 
to add $(\tO^{[\Phi,\Psi]_J}_0+\tO^{[\Phi,\Psi]_J}_J)$ to the set of operators ${\cal O}^{[\Phi,\Psi]_J}_m$
and construct the eigenvectors of $\Delta$. The mixing would be of order $1/J$.  

However, the correspondence between string theory in plane wave background and gauge theory is
supposed to hold in the large $N$ and $J$ limits and this mixing would just make life complicated. 
In support of this last statement, we notice that the expectation value of $\Delta$ for finite $N$ is identical 
to the one for infinite $N$:
\be
\langle{\cal O}^{[\Phi,\Psi]_J}_m\Delta{\cal O}^{[\Phi,\Psi]_J}_m\rangle=
4g_{YM}^2N\left(4\sin^2\pi\varphi_m^J\right)~~.
\ee
This is because the $1/J$ term in the scalar product of operators ${\cal O}^{[\Phi,\Psi]_J}_m$
is canceled by the second term in the equation above. 

This observation suggests that we can take the limit $N\rightarrow\infty$ at the beginning of the 
computations. This prescription solves the problems mentioned above. In particular, the BMN operators
are orthonormal at tree level\footnote{In computing the scalar product of operators, changing the 
summation over $l$ from $0$ to $J$ to an  integral, as in \cite{MITH},
is equivalent to  taking $N\rightarrow\infty$ 
at the beginning of the computation.} and are the right gauge theory duals of the states of the free string theory
to all orders in $\lambda'$ and zero-th order in $g_2$. The equation (\ref{invFt}) can then be modified to a 
more symmetric form
\be
K^{[\Phi,\Psi]_J}\tO^J_l={1\ov J}\sum_{n=0}^{J-1}e^{-2\pi i\,l{n\ov J}}{\cal O}^{[\Phi,\Psi]_J}_n~~~~~
(\forall)~~~l=0\dots J(=\infty)~~.
\label{modinv}
\ee

These subtleties can be avoided by slightly changing the definition of the phase appearing in
equation (3.1) to:
\be
\varphi_m^J={m\over J+1}~~.
\ee
At order ${\cal O}(g_2^0)$ these operators are eigenvectors of the generators of scale 
transformations to all orders in $\lambda'$ and their scalar product becomes diagonal 
to all orders in $1/J$. The expression of $\tO^J_l$ in terms of these modified operators
is obtained from (\ref{modinv}) via obvious replacements. These modifications were also suggested 
in \cite{MODOP}.

An alternative way of performing the computations, which does not require writting $\tO$ in terms of $\cO$,  
goes as follows.  We first compute the contribution to the matrix elements of the Hamiltonian
due to the mixing of the double-trace and single-trace operators induced by the interactions described in the 
previous section. This is the matrix:
\be
H=S\bfg^T~~.
\ee
Then, we perform the change of basis that diagonalizes the scalar product matrix, $S$:
\be
S~~\longrightarrow~~{\hat S}=USU^T=\id
\ee
Under this
transformation, the Hamiltonian matrix transforms as:
\be
H~~\longrightarrow~~{\hat H}=U S\bfg^T U^T~~.
\ee 
Since the new basis is orthonormal, ${\hat H}$ is the appropriate matrix for comparison with string theory.

In the following sections we will adopt the first computational scheme, i.e. we will compute the anomalous 
dimension matrix in the limit $N\rightarrow\infty$. We will then proceed in the next section with the necessary 
change of basis.


\subsection{Scale transformations of double-trace operators}

Having discussed in detail all necessary ingredients, let us now proceed
with the computation of the entries of the anomalous dimension matrix 
describing the appearence of single-trace operators in the scale transformations 
of double-trace ones. Given the qualitative differences between 
${\cal O}_{0,0}^{[\Phi]_{J_1}[\Psi]_{J-J_1}}$ and ${\cal O}_{m,\emptyset}^{[\Phi,\Psi]_{J_1}[\id]_{J-J_1}}$, 
we will treat then separately. For both cases 
the template Feynman diagrams are those in figure \ref{algmFd} with the
empty and filled dots being either a $Z$ or an impurity.

\subsubsection{${\cal O}^{[\Phi]_{J_1}[\Psi]_{J-J_1}}_{00}\longrightarrow {\cal O}^{[\Phi,\Psi]_J}_m$}

$\bullet$ $Z-Z$ interactions:

Each of the diagrams in  figure \ref{algmFd} gives a double sum, since each of the 
two $Z$-s can sit anywhere in the original string of $Z$-s of each operator. The first
two diagrams are, of course, identical. Taking into 
account the relative factors in the first line of figure \ref{divgraphs} we find that
the relevant combination of operators is: 
\bea
\sum_{p=0}^{J_2-1}\sum_{q=0}^{J_1-1}\!\!\!\!\!\!\!
&{}&\left[2Tr[\Psi Z^{p+q+1}\Phi Z^{J-p-q-1}]-Tr[\Psi Z^{p+q+2}\Phi Z^{J-p-q-2}]
\right.\nn\\
&{}&\left.-Tr[\Psi Z^{p+q}\Phi Z^{J-p-q-1}]\right]~~.
\label{ZZ}
\eea
It is clear that most terms cancel and we are left with a counterterm equal to:
\be
\C_1= {1\ov\epsilon}g_{YM}^2\T(\Psi,\Phi)\, ,
\ee
where $\T(\Psi,\Phi)$ is defined to be the result of the sum in (\ref{ZZ}):
\be
\T(\Psi,\Phi)=
Tr[\Psi Z^{J_1}\Phi Z^{J_2}]+Tr[\Psi Z^{J_2}\Phi Z^{J_1}]-
Tr[\Psi \Phi Z^{J_1+J_2}]-Tr[\Phi\Psi Z^{J_1+J_2}]~~.
\ee

$\bullet$ ${\rm impurity}-{\rm impurity}$ interactions:

In this case each diagram in figure \ref{algmFd} gives exactly one term. Taking into 
account the second line of figure \ref{divgraphs}, the required counterterm is:
\be
\C_2= {1\ov\epsilon}g^2_{YM}\T(\Psi,\Phi)\, .
\ee

$\bullet$ $Z-{\rm impurity}$ interactions:

Interactions $\Phi Z$ and $\Psi Z$ are described by the third line
in figure  \ref{divgraphs}  as well as six more diagrams in which $\Psi$ is replaced by 
$\Phi$ and its position is switched with the position of $Z$. 

Each diagram in  figure \ref{algmFd} leads to a simple sum, corresponding to the position 
of the $Z$ field in the original oprator. It is not hard to see that these sums cancel pairwise. 
Indeed, consider the case in which the filled dot in figure \ref{algmFd} represents an impurity, 
say $\Phi$, while the empty one is a $Z$ field. Then, taking into account the relative signs in 
the third line of figure  \ref{divgraphs} we have the following coefficient for ${1\ov \epsilon}$:
\bea
&&\sum_{p=0}^{J_2-1}Tr[\Psi Z^{J_1+p+1}\Phi Z^{J_2-p-1}]
+\sum_{p=0}^{J_1-1}Tr[\Psi Z^{p}\Phi Z^{J-p}]\nn\\
&&-\sum_{p=0}^{J_2-1}Tr[\Psi Z^{J_1+p+1}\Phi Z^{J_2-p-1}]
-\sum_{p=0}^{J_1-1}Tr[\Psi Z^{p}\Phi Z^{J-p}]=0
\eea
and they obviously cancel. Each term in the equation above is associated to one of the four 
diagrams in figure \ref{algmFd}, respectively.

Thus, the total counterterm needed for the renormalization of ${\cal O}_x$ is:
\be
\C={2\ov\epsilon}g_{YM}^2 \T(\Phi,\Psi)~~.
\ee

Using equations (\ref{invFt}) we can express $\T$ in terms of BMN operators:
\be
\T=-{4\ov \sqrt{J}}\sum_{m=0}^{J-1} \sin^2\left(\pi\,m {J_1\ov J}\right){\cal O}^{[\Phi,\Psi]_J}_m
\ee

Thus, the total renormalization constant for the mixing between a double-trace operator
built out of two BPS states and the single-trace BMN operators becomes:
\be
Z_{x}{}^i=-{1\ov\epsilon}\,{8g_{YM}^2\ov \sqrt{J}}\sin^2\left(\pi\,m_i {J_i\ov J}\right)
~~~(\forall)~~m_i=0,\dots,J-1
\ee
Using (\ref{anomdimdef}), this immediately leads to the following entries of the anomalous 
dimension matrix:
\be
\gamma_{x;i}=-{16g_{YM}^2\ov \sqrt{J}}
\sin^2\left(\pi\,m_i {J_2\ov J}\right)
\ee
where in the last equality we used the coefficient of the free 3-point function 
derived in the appendix.


\subsubsection{${\cal O}^{[\Phi,\Psi]_{J_1}[\id]_{J-J_1}}_{n,\emptyset}
\longrightarrow {\cal O}^{[\Phi,\Psi]_J}_m$}

Now we turn to the computation of the admixture of single-trace operators in the scale 
transformation of the double-trace operators ${\cal O}^{[\Phi,\Psi]_{J_1}[\id]_{J-J_1}}_{m,\emptyset}$. 
Since one of the operators
building  ${\cal O}^{[\Phi,\Psi]_{J_1}[\id]_{J-J_1}}_{m,\emptyset}$ does not contain any impurities, 
there are only two types of interactions to analyze: $Z-Z$ and $Z-{\rm impurity}$. 

However, things are somewhat simpler than they seem, since it is easy to see that the 
$Z-Z$ interactions vanish. Indeed, let us consider more general 
operators, of the form $Tr[AZ^p]$ and $Tr[BZ^q]$. Particular choices of $A$ and $B$  lead 
to the operators $\tO_l$. Using the first line in figure \ref{divgraphs}, we find that the divergence
in a 1-loop graph with one insertion of the product of these operators is proportional to: 
\bea
Tr[A Z^{p} B Z^{q}]+Tr[A Z^{q}B Z^{p}]
-Tr[AB Z^{p+q}]-Tr[BA Z^{p+q}]~~.
\eea
For the operators ${\cal O}^{[\Phi,\Psi]_{J_1}[\id]_{J-J_1}}_{m,\emptyset}$
one of the operators $A$ and $B$ is the identity operator 
and thus, the sum above vanishes.

Thus, unlike the case of operator ${\cal O}^{[\Phi]_{J_1}[\Psi]_{J-J_1}}_{0,0}$, 
now the $Z$-impurity interactions give the 
nonvanishing contribution. Taking into account the normalization factors, 
each of the $J+1$ terms of ${\cal O}^{[\Phi,\Psi]_{J_1}[\id]_{J-J_1}}_{m,\emptyset}$  
gives the following counterterms:
\be
\C_l={1\ov\epsilon} {1\ov \sqrt{J_1J_2}}\,2J_2\,\S_l
\label{interm}
\ee
with $\S_l$ defined as
\bea
\S_l&=&Tr[\Phi  Z^l\Psi Z^{J-l}] - Tr[\Phi  Z^{l+1}\Psi Z^{J-l-1}]\nn\\ 
&+&Tr[\Phi  Z^{J_2+l}\Psi Z^{J_1-l}] - Tr[\Phi  Z^{J_2+l-1}\Psi Z^{J_1-l+1}] ~~.
\label{sl}
\eea
Each of the four terms above comes from one of the four diagrans in figure \ref{algmFd}.
The extra factor of $2$ in (\ref{interm}) is due to the fact that both $\Phi$ and $\Psi$ 
bring the same contribution while the extra factor of $J_2$ is due to the $J_2$ equivalent
choices of $Z$ in the $Tr[Z^{J_2}]$ factor of ${\cal O}^{[\Phi,\Psi]_{J_1}[\id]_{J-J_1}}_{m,\emptyset}$.

Let us consider first the case $n=0$, i.e. ${\cal O}^{[\Phi,\Psi]_{J_1}[\id]_{J-J_1}}_{n,\emptyset}$
is a product of BPS operators. 
Then the sum over $l$ simplifies considerably and we find
\be
\sum_{l=0}^{J_1}\S_l=-Tr[\Phi Z^{J_1+1}\Psi Z^{J_2-1} +
\Phi Z^{J_2-1}\Psi Z^{J_1+1}-\Phi\Psi Z^J - \Psi\Phi Z^J]
\label{sumS}
\ee
Using equation (\ref{invFt}) the required counterterms can be expressed in terms 
of BMN operators as:
\be
\C= {8\ov\epsilon}\sqrt{J_2\ov J_1 J} \sum_{m=0}^{J-1}
\sin^2\left(\pi m {J_1+1\ov J}\right) \,{\cal O}^{[\Phi,\Psi]_J}_m
\label{inmassless}
\ee

Therefore, the corresponding entry of the anomalous dimension matrix is:
\be
\gamma_{\alpha=0;i}=16 g_{YM}^2\sqrt{J_2\ov J_1 J} 
\sin^2\left(\pi m {J_1+1\ov J}\right)
\label{ex2rez}
\ee
where in the last equality we neglected terms suppressed by ${1\ov J}$ and
we used the 3-point function coefficient from the appendix.\footnote{In \cite{MITH} this 
subleading term never appears because the sum over $l$ is replaced by an integral. The 
corrections to this replacement are of order ${1\ov J}$}.

We now turn to the case of general ${\cal O}^{[\Phi,\Psi]_{J_1}[\id]_{J-J_1}}_{n,\emptyset}$. 
The sum over $\S_l$ is now:
\bea
\sum_{l=0}^{J_1}e^{il\varphi_n^{J_1}}\S_l&=&\sum_{l=0}^{J_1} \S_l\nn\\
&+& 
\sum_{t=0}^{J_1-1} 
e^{-i t \varphi_n^{J_1}}(1-e^{-i\varphi_n^{J_1}})Tr[\Phi Z^{J_1-t}\Psi Z^{J_2+t}]\nn\\
&+& 
\sum_{t=0}^{J_1-1} 
e^{i t \varphi_n^{J_1}}(1-e^{i\varphi_n^{J_1}})Tr[\Phi Z^{J_2+t}\Psi Z^{J_1-t}]
\label{summassive}
\eea
where $\varphi_n^{J_1}=2\pi {n\ov J_1}$ and we have isolated the terms that do not vanish 
in the limit $n=0$. Using equation (\ref{invFt}) the last two sums can be expressed in terms 
of BMN operators; using that $exp(iJ_1\varphi_n^{J_1})=1$, the result is:
\be
{1\ov \sqrt{J}}\sum_{m=0}^{J-1}\left[\left(e^{-iJ_1\varphi_m^J}-1\right)
{1-e^{-i\varphi_n^{J_1}}\ov 1-e^{-i(\varphi_n^{J_1}-\varphi_m^{J})}}
+
\left(e^{iJ_1\varphi_m^J}-1\right)
{1-e^{i\varphi_n^{J_1}}\ov 1-e^{i(\varphi_n^{J_1}-\varphi_m^{J})}}
\right]{\cal O}^{[\Phi,\Psi]_J}_m~~.
\ee
To leading order in  $1/J$ and $1/J_1$ this equation simplifies to:
\be
{1\ov \sqrt{J}}\sum_{m=0}^{J-1} {n\ov m{J_1\ov J}-n}\,
4\sin^2\left(\pi m {J_1\ov J}\right)
{\cal O}^J_m~~~~~~x={J_1\ov J}~~.
\label{finmassive}
\ee
Therefore, the counterterms required for the renormalization of diagrams with 
one insertion of ${\cal O}^{[\Phi,\Psi]_{J_1}[\id]_{J-J_1}}_{n,\emptyset}$ are
\be
\C_{full}={8 g_{YM}^2\ov \epsilon}\sqrt{J_2\ov J\,J_1} \sum_{m=0}^{J-1}
{mx\ov mx - n}\sin^2\left(\pi m {J_1\ov J}\right){\cal O}^{[\Phi,\Psi]_J}_m
\label{direct}
\ee
which in turn imply that the relevant entries in the anomalous dimension matrix are:
\be
\gamma_{i,\alpha}=16 g_{YM}^2\sqrt{J_2\ov J\,J_1} {m_ix\ov m_ix - n_\alpha}
\sin^2\left(\pi m_i {J_1\ov J}\right)
\ee
where again we have used the 3-point function coefficients from the appendix.


\subsection{Scale transformations of single-trace operators}

To complete the computation of the entries of the anomalous dimension matrix 
describing the mixing of ${\cal O}_m^{[\Phi,\Psi]_J}$ and 
${\cal O}^{[\Phi,\Psi]_{J_1}[\id]_{J-J_1}}_{n,\emptyset}$
under scale transformations
we need to find $\gamma_{A\, i}$. The template Feynman diagrans are those that lead to 
the comultiplication (figure \ref{algcoFd}). As in the previous subsection, we will 
discuss ${\cal O}^{[\Phi]_{J_1}[\Psi]_{J-J_1}}_{0,0}$ and 
${\cal O}^{[\Phi,\Psi]_{J_1}[\id]_{J-J_1}}_{n,\emptyset}$ separately. 

Before we proceed let us stress two points. First, we will be working with an $SU(N)$ 
theory. This will allow us to set to zero terms containing traces of a single field which might
not be in line with the general formulae we will derive. Second, the contribution of the
nearest neighbour interaction leads to the diagonal entries of the anomalous dimension 
matrix and was already discussed in our conventions at the end of section \ref{secdivergences}.
We will however keep those terms as a check for the derivation of more general expressions.

\subsubsection{${\cal O}^{[\Phi,\Psi]_J}_m\longrightarrow {\cal O}^{[\Phi]_{J_1}[\Psi]_{J-J_1}}_{0,0}$}

The mixing turns out to vanish. Indeed, the various interactions generate the following 
counterterms:

$\bullet~~ZZ:$
\be
\C_{ZZ}=-{1\ov\epsilon}{g_{YM}^2\ov\sqrt{J}}\sum_{l=1}^{J-1}e^{il\varphi_m^J}
\left(O^l_\Phi O^{J-l}_\Psi+O^{J-l}_\Phi O^{l}_\Psi\right)
\ee

$\bullet~~\Phi\Psi:$
\be
\C_{\Phi\Psi}=-{1\ov\epsilon}{g_{YM}^2\ov\sqrt{J}}\sum_{l=1}^{J-1}e^{il\varphi_m^J}
\left(O^l_\Phi O^{J-l}_\Psi+O^{J-l}_\Phi O^{l}_\Psi\right)
\ee

$\bullet~~Z\Phi:$
\be
\C_{Z\Phi}={1\ov\epsilon}{g_{YM}^2\ov\sqrt{J}}\sum_{l=1}^{J-1}e^{il\varphi_m^J}
\left(O^l_\Phi O^{J-l}_\Psi+O^{J-l}_\Phi O^{l}_\Psi\right)
\ee
The $Z\Psi$ interactions produce the same result. 

Thus, doubling the $Z\Phi$  and adding to it the $ZZ$ and $\Phi\Psi$ contributions 
we find a vanishing counterterm, as stated above. Therefore, 
\be
\gamma_{i, x}=0
\ee


\subsubsection{${\cal O}^{[\Phi,\Psi]_J}_m\longrightarrow 
{\cal O}^{[\Phi,\Psi]_{J_1}[\id]_{J-J_1}}_{n,\emptyset}$}

For operators of this type the mixing turns out to be generically nonvanishing. Furthermore, 
all types of interactions give nonvanishing contributions. As in the previous subsections, 
most of our derivations are valid at finite $J$ as well. A byproduct of this computation will be 
that at subleading order in ${1\ov J}$ the BMN operators mix with non-BMN ones. 

Let us now analyze the three possible types of interactions:

\noindent
$\bullet~~Z-Z$

The required counterterm is:
\bea
&&\C_{ZZ}=-{1\ov\epsilon}{g_{YM}^2\ov \sqrt{J}}\times\\
&&\left[\sum_{l=2}^J e^{il\varphi_m^J}
\sum_{p=0}^{l-2} (l-p-1)\left[-2 O^{p+1}\tO_{l-p-1}^{J-p-1}+ O^{p}\tO^{J-p}_{l-p}+
O^{p+2}\tO^{J-p-2}_{l-p-2}\right]\right.\nn\\
&+&\left.\sum_{l=0}^{J-2}e^{il\varphi_m^J}
\sum_{p=0}^{l-2} (l-p-1)\left[-2 O^{p+1}\tO_{l}^{J-p-1}+ O^{p}\tO^{J-p}_{l}+
O^{p+2}\tO^{J-p-2}_{l}\right]\right]\nn
\eea
where $O^p=Tr[Z^p]$. One of the sums can be done explicitly, with the result:
\bea
\C_{ZZ}&=&-{1\ov\epsilon}{g_{YM}^2\ov \sqrt{J}}
\sum_{l=2}^J 
\left[e^{il\varphi_m^J}\tO^{J-l}_{0}+e^{-il\varphi_m^J}\tO^{J-l}_{J-l}\right]\,O^l\nn\\
&&-{1\ov\epsilon}{g_{YM}^2N\ov \sqrt{J}}\left[(J+2)-4\right] 
\sum_{l=2}^J e^{il\varphi_m^J}\tO_l^J
\eea
The second line in the formula above is due to the nearest neighbour interaction
and is one of the terms leading to the anomalous dimension of ${\cal O}^{[\Phi,\Psi]_J}_{m}$. 
As stated in the beginning, we  kept it in place as a check of our computation and it 
is reassuring that its coefficient is the correct one: $J+2$ is canceled by the individual 
wave function renormalization of fields while the remaining $(-4)$ is twice as large as 
the $Z-Z$ contribution to equation (\ref{anomdim}). 

\

\noindent
$\bullet~~\Phi-\Psi$

There are two diagrams that give nonvanishing contributions to the mixing between 
${\cal O}_m^{[\Phi,\Psi]_J}$ and ${\cal O}^{[\Phi,\Psi]_{J_1}[\id]_{J-J_1}}_{n,\emptyset}$. 
They correspond to the second and third columns of 
the second line of figure \ref{divgraphs}. It is not hard to see that the required counterterm is:
\be
\C_{\Phi\Psi}=-{1\ov\epsilon}{g_{YM}^2\ov \sqrt{J}}\sum_{l=0}^J
\left[e^{il\varphi_m^J}\tO^{J-l}_{J-l}+e^{-il\varphi_m^J}\tO^{J-l}_{0}\right]\,O^l
\ee

\

\noindent
$\bullet~~Z\Phi~~{\rm and}~~Z\Psi$

As one probably expects, each type of interaction requires the same counterterm. 
Carefully following the diagrams in figure \ref{algcoFd}, the counterterms turn
out to be:
\bea
\C_{Z\Phi}=-{1\ov \epsilon} {g_{YM}^2\ov\sqrt{J}}&&\!\!\!\!\!\!\!
\left[~~
\sum_{l=1}^Je^{il\varphi_m^J}\sum_{p=0}^{l-1}\left[O^p\tO^{J-p}_{l-p-1}
-O^{p+1}\tO^{J-p-1}_{l-p-1}\right]\right.\nn\\
&&\left.
+\sum_{l=0}^{J-1}e^{il\varphi_m^J}\sum_{p=0}^{J-l-1}\left[O^p\tO^{l+1}_{l-p-1}
-O^{p+1}\tO^{J-p-1}_{l}\right]\right]
\eea
This equation can be simplified to:
\bea
\C_{Z-\Phi}=-{1\ov \epsilon} {g_{YM}^2\ov\sqrt{J}}&&\!\!\!\!\!\!\!
\left\{~~
\sum_{t=0}^{J-1}\left[ ~~e^{-it\varphi_m^J}\sum_{p=1}^{J-t-1}O^p\left[\tO^{J-p}_{J-t-p-1}
-\tO^{J-p}_{J-t-p}\right]\right.\right.\nn\\
&&~~~~\left.+e^{it\varphi_m^J}\sum_{p=1}^{J-t-1}O^p\left[\tO^{J-p}_{t+1}
-\tO^{J-p}_{t}\right]
\right]\nn\\
&&\!\!\!
-\sum_{l=1}^J\,O^l\left[e^{il\varphi_m^J}\tO^{J-l}_{0}
+e^{-il\varphi_m^J}\tO^{J-l}_{J-l}\right]
\nn\\
&&\!\!\!\left.
+N\sum_{l=1}^Je^{il\varphi_m^J}\tO^{J}_{l-1}+N\sum_{l=0}^{J-1}
e^{il\varphi_m^J}\tO^{J}_{l+1}
\right\}
\label{ZPhi}
\eea
As in the case of $ZZ$ interactions, the terms on the last line are the source of 
anomalous dimension for the operator ${\cal O}^{[\Phi,\Psi]_J}_{m}$. Taking into account the 
contribution of both $\Phi$ and $\Psi$ (i.e. doubling (\ref{ZPhi})) we indeed find the 
correct coefficient (equations (\ref{anomdim}) and (\ref{fullanomdim})).

To extract the desired entries of the anomalous dimension matrix we isolate
double-trace operators containing $O^{J_2} = Tr[Z^{J-J_1}]$. The formulae simplify 
considerably:
\be
\C_{ZZ}+\C_{\Phi\Psi}=-{1\ov\epsilon}{g_{YM}^2\ov \sqrt{J}}
\left(e^{iJ_1 \varphi_m^J}+e^{-iJ_1\varphi_m^J}\right)\left[\tO^{J_1}_{0}+
\tO^{J_1}_{J_1}\right]\,O^{J_2}
\ee

When isolating  $O^{J_2}$ from $Z_{Z\Phi}$ we have to be careful with the upper 
limit for the summation over $t$, since we want $1\le p=J_2\le J- t - 1$. Taking this properly
into account, the result is: 
\bea
\C_{Z\Phi}=-{1\ov\epsilon}{g_{YM}^2\ov \sqrt{J}}&&\!\!\!\!\!\!\!
\left\{
\sum_{t=1}^{J_1-1}\left[e^{-it\varphi_m^J}(e^{i\varphi_m^J}-1)\tO_{J_1-t}^{J_1}+
e^{it\varphi_m^J}(e^{-i\varphi_m^J}-1)\tO_{t}^{J_1}\right]\,O^{J_2}\right.\nn\\
&&+e^{-it\varphi_m^J}(e^{i\varphi_m^J}-1)\tO_{0}^{J_1}O^{J_2}+
e^{it\varphi_m^J}(e^{-i\varphi_m^J}-1)\tO_{J_1}^{J_1}O^{J_2}\nn\\
&&-\left[\tO^{J_1}_{0}+\tO^{J_1}_{J_1}\right]\,O^{J_2}
\eea

Combining everything we find that the counterterm required for the diagrams
with one ${\cal O}_i$ insertion which produce a double-trace operator containing
$O^{J_2}$ is:
\bea
\C^{split}=-{1\ov\epsilon}{g_{YM}^2\ov \sqrt{J}}&&\!\!\!\!\!\!\!
\left\{
2\sum_{t=1}^{J_1-1}\left[e^{-it\varphi_m^J}(e^{i\varphi_m^J}-1)\tO_{J_1-t}^{J_1}+
e^{it\varphi_m^J}(e^{-i\varphi_m^J}-1)\tO_{t}^{J_1}\right]\,O^{J_2}\right.\nn\\
&&+2\left[e^{-it\varphi_m^J}(e^{i\varphi_m^J}-1)\tO_{0}^{J_1}+
e^{it\varphi_m^J}(e^{-i\varphi_m^J}-1)\tO_{J_1}^{J_1}\right]O^{J_2}\nn\\
&&+\left(e^{iJ_1 \varphi_m^J}+e^{-iJ_1\varphi_m^J}-2\right)
\left[\tO^{J_1}_{0}+\tO^{J_1}_{J_1}\right]\,O^{J_2}
\eea

It is now a simple exercise to use equation (\ref{invFt}) and compute the sum over 
$t$ we find to express the $\C^{split}$ in terms of BMN operators:
\bea
\C^{split}=-{1\ov\epsilon}{g_{YM}^2\ov \sqrt{J \,J_1}}&&\!\!\!\!\!\!\!
\sum_{n=0}^{J_1-1}    {\cal O}^{[\Phi,\Psi]_{J_1}[\id]_{J-J_1}}_{n,\emptyset}
\left\{-8\sin^2\left({1\ov 2}J_1 \varphi_m^J\right)+\right.\\
&&\left.
\!\!\!\!\!\!\!\!\!\!\!\!\!\!\!\!\!\!\!\!\!\!\!\!\!\!
\!\!\!
+2\left[
(1-e^{-iJ_1\varphi_m^J})
{e^{i\varphi_m^J}-1\ov e^{i(\varphi_m^J-\varphi_n^{J_1})}-1}+
(1-e^{iJ_1\varphi_m^J})
{e^{-i\varphi_m^J}-1\ov e^{-i(\varphi_m^J-\varphi_n^{J_1})}-1}
\right]\right\}\nn\\
-{1\ov\epsilon}{g_{YM}^2\ov \sqrt{J }}~~&&\!\!\!\!\!\!\!\!\!\!\!\!
\left\{e^{-it\varphi_m^J}(e^{i\varphi_m^J}-1)
-e^{it\varphi_m^J}(e^{-i\varphi_m^J}-1)\right\}\left[\tO_{0}^{J_1}
-\tO_{J_1}^{J_1}\right]O^{J_2}~~.
\nn
\eea
Expanding this for large $J_i$ we find:
\be
\C^{split}=-{8 g_{YM}^2\ov\epsilon}\sqrt{J_2\ov J\,J_1} \sum_{m=0}^{J-1}
{n\ov mx - n}\sin^2\left(\pi m {J_1\ov J}\right){\cal O}^{[\Phi,\Psi]_{J_1}[\id]_{J-J_1}}_{n,\emptyset}
\label{conjugate}
\ee 
which clearly leads to the following entries in the anomalous dimension matrix:
\bea
\gamma_{\alpha,i}=-16 g_{YM}^2\sqrt{J_2\ov J\,J_1} {n\ov mx - n}
\sin^2\left(\pi m {J_1\ov J}\right)
\eea


\subsection{Symmetry}

An essential test for identification of the Hamiltonian is its Hermiticity.
Since the scalar product is not diagonal we use equation (\ref{matelem}) to calculate its 
matrix elements:
\be
H_{iA}=S_{i\Sigma}\gamma_{A\Sigma}=\gamma_{Ai}+C_{iA}\delta_A~~~~~~~~
H_{Ai}=S_{A\Sigma}\gamma_{i\Sigma}=\gamma_{iA}+C_{iA}\delta_i
\ee
Thus we find that the Hamiltonian is Hermitian if
\bea
\gamma_{A i} - \gamma_{iA}=C_{iA}(\delta_i-\delta_A)~~.
\label{check1}
\eea

It is not hard to check that the anomaloud dimension matrix satisfies this requirement.
To summarize the results of this long computation, we list here all entries of the
anomalous dimension matrix:
%
\bea
\bfg&\equiv& \pmatrix{\gamma_{ij}&\gamma_{iA}\cr
\gamma_{B j}& \gamma_{BA}\cr}\\
&=&16\pi^2\lambda' \pmatrix{ 
{m^2_i}\delta_{ij}& 0& -{g_2\over \pi^2\sqrt{J}} \sqrt{1-x\over x}{n_\alpha\over
m_i x-n_\alpha}\sin^2\pi x m_i \cr
-{g_2\over \pi^2\sqrt{J}}\sin^2\pi x m_i& 0& 0\cr
{g_2\over \pi^2\sqrt{J}} \sqrt{1-x\over x}{m_ix\over
m_i x-n_\alpha}\sin^2\pi x m_i & 0 &n_\alpha\delta_{\beta\alpha}\cr}\nn
\label{gamma}
\eea
Inspecting (\ref{gamma}) and using the explicit expressions for $\delta$ and $C$ it is easy to see
that this equation is satisfied. Thus, at this level,
the generator of scale transformations is hermitian and its interpretation as the string 
Hamiltonian the interpretation is reasonable. This can be also interpreted as a test of the 
correctness of our computation.


\section{Diagonalization of the scalar product}

The scalar product of the states should be invariant under
evolution, which is equivalent to the Hamiltonian being symmetric.
From the CFT point of view, this implies that in the basis where
the matrix of the scalar products is the unit matrix, the matrix
of the anomalous dimensions should be symmetric.

To extract the Hamiltonian in a basis suitable for comparison with string theory 
we must diagonalize the scalar product of states. On the string theory side the 
2-string states belong to a space which is the tensor product of 2 one-string states.
This space is orthogonal to all orders in string coupling to the space of 1-string states;
the transition from one to the other is given by the Hamiltonian.

\subsection{Order $g_2$; three-string vertex}

As discussed in section 2, the gauge theory version of the string scalar product 
is defined by the 2-point function. Using the results in the appendix, we have the following 
matrix representation:
\be
S\equiv\pmatrix{S_{ij} & S_{iA}\cr S_{B j}& S_{BA}\cr}=
\pmatrix{\delta_{ij} & C_{iA}\cr C_{jB}& \delta_{BA}\cr}
\equiv
\pmatrix{\id & g_2 c\cr g_2c^T & \id\cr}~~.
\ee
where for later convenience we exibited the $g_2$ dependence of $C$.
This equation is correct to leading order in $g_2\sim g_s$. The diagonal
entries are corrected at order $g_2^2$ \cite{MITH}, \cite{SEST}, while the scalar product
between single- and double-trace operators is corrected at order $g_2^3$. Taking into 
account  all corrections of order $g_2^2$ is the subject of the next subsection.

Let us now  find a basis in which the scalar product is diagonal and the matrix elements 
of the generator of scale transformations match with the string theory predictions for the 
matrix elements of $P^-$. To this end let us introduce the operators ${\hat {\cal O}}$ 
related to $\cO$ by some transformation matrix $U$:
\be
{\hat {\cal O}}_\Sigma=\sum_{\Pi}U_{\Sigma\Pi}\cO_\Pi~~~~~~~
{\rm with}
~~~~~~ ~~~
U=\pmatrix{\delta_{ij} & g_2a_{i A}\cr g_2b_{B j} & \delta_{AB}\cr}
\label{ansU}
\ee
where for later convenience we exhibited the  $g_2$ of various entries. In this basis, the 
scalar product matrix is:
\be
{\hat S}=USU^T
\ee
which equals to the identity matrix up to corrections of order $g_2^2$ provided that
\be
c+a+b^T=0
\label{diagconstraint}
\ee

Performing the transformation (\ref{ansU}) on equation  (\ref{scale}) we find that  
the anomalous dimension matrix expressed in the basis of operators ${\hat O}$ is 
related to $\gamma$ by:
\be
{\hat \bfg}=U\bfg U^{-1}
\label{transfgamma}
\ee
It is clear that while $\gamma$ is not symmetric, ${\hat \gamma}$ can be since
$U$ need not be an orthogonal matrix. Introducing the following notation for $\gamma$
\be
\bfg=\pmatrix{\Delta_1 & g_2\Delta\cr g_2\Delta' & \Delta_2\cr}
\ee
we find that, up to corrections of order $g_2^2$, ${\hat \gamma}$ is given by:
\be
{\hat \gamma}=\pmatrix{\Delta_1 & g_2\left[\Delta-(\Delta_1 a - a\Delta_2)\right]\cr 
g_2\left[\Delta' +(b\Delta_1 - \Delta_2b)\right]& \Delta_2\cr}
\ee
It is easy to see that this matrix is symmetric provided the scalar product is the identity 
matrix. Indeed, 
\bea
&&[\Delta-(\Delta_1 a - a\Delta_2)]-[\Delta' +(b\Delta_1 - \Delta_2 b)]=\\
&=&\Delta-\Delta' -[\Delta_1(a+b^T)  - \Delta_2 (a+b^T) ]\nn
=
\Delta-\Delta'  +[\Delta_1 c - \Delta_2 c ]
\eea
which is satisfied due to (\ref{check1}).

To find an explicit form for ${\hat\bfg}$ we must solve (\ref{diagconstraint}). 
Clearly there is no unique solution. However, it is possible to show that all solutions 
to this equation are related by left-multiplication with a matrix that is orthogonal to  
order $O(g_2^2)$. To this end let us consider some fixed solution $a_0$ and $b_0$ to 
the equation above. Any other solution is related to this one
by
\be
a_0\rightarrow a=a_0+O_{12}~~~~~~~~b_0\rightarrow b=b_0-O_{12}^T~~.
\ee
This follows trivially by subtracting equation (\ref{diagconstraint}) written for
$a$ and $b$ from the one written for $a_0$ and $b_0$. 

Consider now an orthogonal matrix $O$
\be
O=\pmatrix{O_{11} & O_{12}\cr {O_{21}} & O_{22}\cr}~~~~~
O_{11}O_{11}^T+O_{12}O_{12}^T=1~~~~O_{22}O_{22}^T+O_{21}O_{21}^T=1~~~~
O_{11}O_{21}^T+O_{12}O_{22}^T=0~~.
\ee
and construct $OU_0$.
\be
U=OU_0=\pmatrix{O_{11}+g_2O_{12}b & g_2O_{11}a+O_{12} \cr g_2O_{22}b+O_{21}& O_{22}+
g_2O_{21}a\cr}~~.
\ee
For $U$ to have the same form as (\ref{ansU}) we must choose $O_{12}$ and $O_{21}$ to be
of order $O(g_2)$. This implies that up to corrections of order $O(g_2^2)$ we have
\be
O_{11}=\id~~~~~~~~O_{22}=\id~~~~~~~~O_{21}=-O_{12}^T
\ee
and thus $U$ becomes:
\be
U=\pmatrix{\id & g_2a_0+O_{12} \cr g_2b_0-O_{12}^T&\id\cr}~~.
\ee
Thus, all matrices of type (\ref{ansU}) are related by 
left-multiplication with orthogonal matrices, up to $O(g_2^2)$. 

The most general solutions of  equation (\ref{diagconstraint}) is:
\bea
a_{iA}=\rho_{iA}c_{iA}+e_{iA}&&~~~~~~~~~~~~
b_{A i}=\sigma_{iA} c_{iA}-e_{iA}\nn\cr
&&\rho_{iA}+\sigma_{iA}+1=0
\eea
As stressed before, to pick a particular solution
we require that ${\hat\gamma}_{iA}$ matches the string theory predictions for the matrix elements
of the Hamiltonian between a 1- and a 2-string state.

As before, due to the qualitative differences between $\cO_{0,0}^{[\Phi]_{J_1}[\Psi]_{J-J_1}}$ and 
$\cO^{[\Phi,\Psi]_{J_1}[\id]_{J-J_1}}_{n,\emptyset}$,
it is necessary to treat ${\hat\gamma}_{ix}$, ${\hat\gamma}_{i\alpha}$ separately.
\bea
{\hat\gamma}_{ix}&=&0-\delta_i  (\rho_{ix} C_{ix}+e_{ix})\nn\\
{\hat\gamma}_{i\alpha}&=&
-{\lambda'g_2\over \pi^2\sqrt{J}} \sqrt{1-x\over x}{n_\alpha\over
m_i x-n_\alpha}\sin^2\pi x m_i 
-(\delta_i-\delta_\alpha)(\rho_{i\alpha} C_{i\alpha}+e_{i\alpha})
\eea

We can now choose $\rho$ and $e$ such that ${\hat\gamma}_{ix}$ and 
${\hat\gamma}_{i\alpha}$ match the string theory predictions \cite{SPVOIIv3}. 
As discussed above
this is not a unique choice, but all such choices are related by orthogonal 
transformations. 
\be
\rho_{iA}=\sigma_{iA}=-{1\ov 2}~~~~~~~~~~~~e=0
\ee
It is easy to see that all constraints are satisfied and the anomalous dimension 
matrix is:
\be
{\hat \bfg}=\pmatrix{16\pi^2\lambda' m_i\delta_{ij} & -{\lambda'g_2\over 2\pi^2\sqrt{J}}\sin^2\pi x m_i &
{\lambda'g_2\over 2\pi^2\sqrt{J}}
\sqrt{1-x\over x}\sin^2\pi x m\cr
 -{\lambda'g_2\over 2\pi^2\sqrt{J}}\sin^2\pi x m_j& 0&0\cr 
{\lambda'g_2\over 2\pi^2\sqrt{J}}
\sqrt{1-x\over x}\sin^2\pi x m
&0& 16\pi^2\lambda' n_\alpha\delta_{\beta\alpha}\cr}~~.
\ee
Since the scalar product is diagonal, ${\hat\gamma}$ is the matrix representation 
of the generator of scale transformations and it matches the string theory prediction for the Hamiltonian matrix. 
For later convenience let us list $A$\footnote{Mixing of operators was discussed in \cite{SESTII}
as well as in \cite{STRINGS}. The precise formulae presented there are different from these ones. 
Our operator redefinitions seem to be  related to those of \cite{SESTII} by an orthogonal transformation 
of the type described earlier in this section.}:
\bea
a_{iA}=b_{Ai}=-{1\ov 2}c_{iA}
\label{U0}
\eea
as well as the operator basis dual to string states at order $g_2$:
\bea
{\hat \cO}_i=\cO_i-{1\over 2}\sum_{A}C_{iA}\cO_A~~~~{\hat \cO}_A=\cO_A-{1\over 2}\sum_{i}C_{iA}\cO_i~~.
\label{newoperators}
\eea

We have therefore succeded in reconstructing the string theory hamiltonian from
pure gauge theoretic considerations. We identified the operator basis in which the matrix 
elements computed in gauge theory match the string theory computations in the oscillator 
number basis. This identification is not to be regarded as a prediction of gauge theory, since it was specifically
constricted by comparing the matrix elements of the Hamiltonian. However, now that the state-operator map
is known to order $g_2$, all matrix elements between 1- and 2-string states of all operators should 
match between string and gauge theory. The discussion is, however, not finished yet.


\subsection{Order $g_2^2$; the state - YM operator map; contact terms}

In the previous section we found an operator redefinition which brought the
anomalous dimension matrix to a symmetric form in the sector describing 
the mixing between operators dual to one and two-string states. However, analyzing 
the higher order corrections to $\bfg$ we are led to the conclusion that the anomalous 
dimension matrix should be symmetric to order $g_2^2$. Indeed, $\gamma_{iA}$ 
receives corrections only at order $g_2^3$ while $\gamma_{ij}$ receives corrections 
already at order $g_2^2$. The diagonalization discussed in the previous section also
contributes terms of order $g_2^2$ to ${\hat\gamma}_{ij}$ and ${\hat\gamma}_{BA}$.
These contributions are essentially fixed once the matrix $U_0$ is chosen. 
The corrections of order $g_2^2$ to $\gamma_{ij}$ were computed in detail in \cite{MITH} and 
\cite{SEST}. Since the anomalous dimension matrix transforms as in (\ref{transfgamma}) and 
$\gamma_{ij}$ is already symmetric, it is a nontrivial check that ${\hat\gamma}_{ij}$ 
remains symmetric.

For the scalar product the situation is similar: the corrections to the 2-point
functions of single- and double-trace operators are of order $O(g_2^3)$ while
the corrections to the 2-point functions of single-trace operators are of order $O(g_2^2)$ 
and were computed in \cite{MITH} and \cite{SEST}. 

Since to this order $C_{iA}$ is not modified at order $g_2^2$, it follows that the only modification 
that we are allowed to do to the previous analysis is to add terms of  order $g_2^2$ to 
$U_{ij}$ and $U_{BA}$. These additions are nevertheless uniquely fixed by
the requirement that the scalar product be diagonal to order $O(g_2^2)$.

Let us now proceed with the computation. Using the results of \cite{MITH} and \cite{SEST}
the scalar product is:
\be
S=\pmatrix{\delta_{ij}+g_2^2\mu_{ij} & g_2c_{iA}\cr
                        g_2c_{jB}&          \delta_{BA}+g_2^2 M_{BA}\cr}
\ee
with 
\be
\mu_{ij}=\left\{\begin{array}{rrrcc}
{1\ov 60}+{21 - 2\pi^2m_i^2\ov 48 \pi^4 m_i^4}  & m_i=m_j\ne 0\\
{105 + 8\pi^2 m_i^2\ov 384 \pi^4 m_i^4} & m_i=-m_j\ne 0\\
{2\pi^2 (m_i-m_j)^2-3\ov 24\pi^4 (m_i-m_j)^4}+
{2m_i^2-3m_im_j+2m_j^2\ov 8\pi^4m_i^2m_j^2(m_i-m_j)^2}& |m_i|\ne |m_j|\ne 0
\end{array}\right.
\ee
(we list only the cases when the single-trace operators are non-BPS since in the other 
cases the mixing with double-trace operators vanishes).
Here we have made explicit the fact that the scalar product between single- and 
double-trace operators is of order $g_2$:  $C_{iA}=g_2c_{iA}$. The matrix $M_{BA}$ has not been 
computed. We will determine it by requiring that $\hat\bfg$ be symmetric in this 
sector as well. Then, modifying $U_{ij}$ and $U_{BA}$ with terms of order $g_2^2$, 
the matrix $U$ and its inverse become:
\be
U=\pmatrix{1+g_2^2 d & g_2\,a\cr g_2\,b& 1+g_2^2 {\hat d}\cr}~~;~~
U^{-1}=\pmatrix{1-g_2^2 (d -ab)& -g_2\,a\cr -g_2\,b& 1-g_2^2 ({\hat d}-ba)\cr}
\ee
From the requirement that $U$ diagonalizes $S$ to order $g_2^2$ we find, besides 
equation (\ref{diagconstraint}),  two new constraints:
\bea
\mu+d+d^T+ac^T+ca^T+aa^T&=&0\nn\\
M+{\hat d}+{\hat d}^T+bc+ c^Tb^T+bb^T&=&0
\label{extraconstraints}
\eea
Let us analyze the consequences of these relations.

The first equation above  uniquely determines the symmetric part of $d$ (up to ambiguities 
described in the previous section). We find:
\bea
(d+d^T)_{ij}&=&-\mu_{ij}+{3\ov 4}\sum_A c_{iA}c_{jA}={1\ov 2}\mu_{ij}~~.
\label{symd}
\eea
The sum over $A$ represents a summation over the discrete index labeling the doulble-trace opeartors 
$n_A$ as well as an integration over ${J_1\ov J}=x\in[0,\,1]$.

Having determined the symmetric part of $d$ we can now study ${\hat\gamma}_{ij}$
to order $g_2^2$\footnote{${\hat\gamma}_{iA}$ is unchanged to order $g_2^2$ compared 
with the previous section.}. A trivial computation reveals that
\bea
\label{problem}
{\hat\gamma}_{ij}\!\!&=&\!\!(\Delta_1)_{ji}
+g_2^2[(\delta_j-\delta_i)d_{ij}+(\delta_i-\delta_A)a_{iA}b_{Aj}+
a_{iA}{\Delta}'{}_{Aj} -{\Delta}_{iA}b_{Aj}]\\
{\hat\gamma}_{AB}\!\!&=&\!\!(\Delta_2)_{AB}
+g_2^2[(\delta_B-\delta_A){\hat d}_{AB}+(\delta_A-\delta_i)b_{Ai}a_{iB}+
b_{Ai}{\Delta}_{iB}-{\Delta}'{}_{Ai}a_{iB}]\nn
\eea
In the first expression above a sum over $A$ is implied as well as an integral
over $x\in[0,1]$. These last expressions have a smooth 
limit $g_2\rightarrow 0$,  since both $\Delta$ and $\Delta'$ are proportional to $g_2$. Also, 
$(\Delta_1)_{ij}$ and $(\Delta_2)_{ij}$ contain the order $g_2^0$ diagonal part
$\delta_i\delta_{ij}$ and $\delta_A\delta_{AB}$  together with order $g_2^2$ 
entries. In the second line above a sum over $i$ over all integers is implied. 

Let us analyze in detail the terms of order $g_2^2$ in ${\hat\gamma}_{ij}$. We have 
to make sure that these terms lead to a symmetric anomalous dimension matrix, 
otherwise its interpretation as Hamiltonian is not valid. This might present a problem, 
since (\ref{problem}) is certainly not manifestly symmetric. 

We can proceed in various ways, including brute force computation, but it is more 
instructive to isolate the symmetric and antisymmetric parts of ${\hat\gamma}_{ij}$.
We will use for $a$ and $b$ the expressions
computed in the previous section, equation (\ref{U0}). It is trivial to see that the freedom of 
multiplying $U_0$ from the left by an orthogonal matrix survives at order $g_2^2$. Indeed, 
if ${\hat\bfg}=U_0\bfg U_0^{-1}$ were symmetric, then so is 
${\hat \bfg}_O=O{\hat\bfg}O^T$.

We will write explicitly the antisymmetric part, leaving for the time being the 
symmetric part in terms of the matrix $a$. With these clarifications, 
contributions of order $g_2^2$ to ${\hat\gamma}_{ij}$ due to the diagonalization of 
the scalar product are:
\bea
{\hat \gamma}_{ij}\Big|_{g_2^2}&=&-{1\ov 2}(\delta_i-\delta_j)
(d+d^T+ac^T+ca^T+aa^T)_{ij} + {1\ov 2}(\Delta_{iA}c_{jA}-\Delta_{jA}c_{iA})\nn\\
&-&{1\ov 2}(\delta_i+\delta_j-2\delta_A)a_{iA}a_{jA}
+(\delta_j-\delta_i)(a_{iA}c_{jA}
-a_{jA}c_{iA})\\
&&~~~~~~~~
+\Delta_{iA}({1\ov 2}c_{jA}+a_{jA})+\Delta_{jA}({1\ov 2}c_{iA}+a_{iA})
+{1\over 2}(\delta_i-\delta_j)(d_{ij}-d_{ji})\nn
\eea
with the same provisions for the summation over $A$.

Let us analyze the possibly troublesome first line in the equation above. Using equation
(\ref{extraconstraints}) as well as the explicit expression for $\Delta_{iA}$ from equation
(\ref{gamma}) we find that it becomes:
\be
{1\ov 2}(\delta_i-\delta_j)\mu_{ij}-8\pi^2 \lambda' g_2 (m_i-m_j) \sum_n {n_\alpha\over x}
c_{i\alpha}c_{j\alpha}=0
\ee
Thus, the potential problematic antisymmetric contribution to the anomalous dimension
matrix vanishes.
Thus, we are therefore left with:
\bea
{\hat\gamma}_{ij}\Big|_{g_2^2}&=&-{1\ov 2}(\delta_i+\delta_j-2\delta_A)a_{iA}a_{jA}
+(\delta_j-\delta_i)(a_{iA}c_{jA}
-a_{jA}c_{iA})\nn\\
&&
+\Delta_{iA}({1\ov 2}c_{jA}+a_{jA})+\Delta_{jA}({1\ov 2}c_{iA}+a_{iA})
+{1\over 2}(\delta_i-\delta_j)(d_{ij}-d_{ji})
\label{finalgammahat}
\eea
The sum over $A$ and the integral over $x$ can be easily computed and the result is:
\be
{\hat\gamma}_{ij}\Big|_{g_2^2}={1\over 2}(\delta_i-\delta_j)(d_{ij}-d_{ji})+B_{ij}
\ee
with $B_{ij}$ given by:
\be
B_{ij}=\left\{\begin{array}{cccll}0&i=0~~{\rm or}~~j=0\\
{1\ov 3}+{5\over 2\pi^2 m_i^2}& i=j\ne 0\\
-{15\ov 8\pi^2 m_i^2}&m_i=-m_j\ne0\\
{3\ov 2\pi^2m_im_j}+{1\ov 2\pi^2 (m_i-m_j)^2}&m_i\ne \pm m_j\ne 0
\end{array}\right.
\ee

We interpret this as the contribution of the the contact terms on the string theory side
to the matrix elements of the matrix elements of the Hamiltonian between 1-string states.

The antisymmetric part of $d$ is not fixed from the diagonalization of the scalar 
product. For $|m_i|\ne |m_j|$ we can define it such that
the terms proportional to $g_2^2$ take any value, in particular it can be fixed such that
the matrix elements of the Hamiltonian between operators corresponding to 1-string states
indeed matches the contribution of the contact terms, whatever that may be. This is however 
not the case if $m_i= \pm m_j$ since this implies that $(\delta_i-\delta_j)=0$. Thus, our
analysis predicts that the contribution of the string theory contact terms to the matrix elements 
of the Hamiltonian between 1-string states with equal masses is:
\bea
\langle m|H_{contact}|m\rangle~~&=&{1\ov 3}+{5\over 2\pi^2 m_i^2}\nn\\
\langle m|H_{contact}|-m\rangle&=&-{15\ov 8\pi^2 m_i^2}
\label{contact}
\eea

Now we return to ${\hat\gamma}_{AB}$ and the second equation (\ref{extraconstraints}). 
As in the case of ${\hat\gamma}_{ij}$
everything is in principle determined. While the matrix $M$ describing the
2-point functions of double-trace operators is not known, consistency of our 
procedure determines it: we will require that the antisymmetric part of 
 ${\hat\gamma}_{AB}$ cancels and from here we will find $M_{AB}$.

Using the results of the previous section we find for ${\hat\gamma}_{AB}$:
\bea
{\hat\gamma}_{AB}\Big|_{g_2^2}&=&{1\ov 2}(\delta_A-\delta_B)
M_{AB}+{1\ov 2}(\Delta_{iA}c_{iB}-
\Delta_{iB}c_{iA})\nn\\
&-&{1\ov 8}(\delta_A+\delta_B-2\delta_i)
c_{iA}c_{iB}-{1\ov 2}(\delta_A-\delta_B)({\hat d}_{AB}-{\hat d}_{BA})
\label{22}
\eea
where we have separated the antisymmetric (first line) and symmetric parts (second line) and we 
have already used equation (\ref{extraconstraints}) to eliminate the symmetric part of ${\hat d}$.

Requiring that the antisymmetric part of ${\hat \bfg}$ vanishes
determines all but the diagonal entries of $M$ and the entries corresponding to at least one 
operator being a product of BPS operators. It is not hard to check 
that this equation is consistent for these cases as well. Consider for example 
$\delta_A=\delta_B$. In this case the first 
term manifestly vanishes while the second term vanishes only after the sum over $i$ is 
performed. Making everything explicit we find the following expression for $M$:
\be
M_{AB}=J{g_2^2\over \pi^4} \sqrt{x_A^{3}x_B^{3}(1-x_A)(1-x_B)} 
\sum_{m_i=-J/2}^{J/2}{\sin^2\pi m_i x_A\ov (m_i-{n_A\ov x_A})^2}
{\sin^2\pi m_i x_B\ov (m_i-{n_B\ov x_B})^2}\left[{m_i\over {n_A\ov x_A}
+{n_B\ov x_B}}-1\right]
\label{2pfdouble}
\ee

As for single-trace operators, we can choose the antisymmetric part of 
${\hat d}$ such that almost all off-diagonal entries of ${\hat\gamma}_{AB}$ vanish. 
Indeed, if $(\delta_A-\delta_B)\ne 0$ we can freely set to $0$ the first term in (\ref{22}) 
together with the corresponding entries of $\Delta_2$.
Leaving aside the case $A=-B$ (i.e. $n_\alpha=-n_\beta$) for which a separate 
computation must be done (and we will not do it here), we find that 
${\hat\gamma}_{AA}$ is given by:
\bea
{\hat\gamma}_{AA}=(\Delta_2)_{AA}+{g_2^2\over 4}\sum_{i}
(\delta_i-\delta_A)c_{iA}c_{iA}
\eea
where we explicitly wrote the sum over $i$.

Now we analyze this case by case, i.e. we split the index $A$ into $x$ and $\alpha$. 
Furthermore, the sum over $i$ is rather difficult to perform.
Since the summand is well-behaved at all points, we can replace the sum over $i$ from
$-J/2$ to $J/2$ by an integral over $z\in [-1/2,\, 1/2]$ and multiply by another factor 
of $J$. While this operation is valid up to corrections of order $1/J$, this approximation 
is enough to illustrate that the sum is subleading in $1/J$.
We find:
\bea
{\hat\gamma}_{xx}&=&(\Delta_2)_{xx}+{g_2^2\over 4}\sum_{i}\delta_ic_{ix}c_{ix}\nn\\
&=&(\Delta_2)_{xx}+{4\ov J\pi^2} \lambda\,g_2^2\sum_{m_i=-J/2}^{J/2}\,{m_i^2\ov J^2}\,
{\sin^4\pi {m_i\ov J}J_1\ov {m^4\ov J^4}} {1\ov J^4}\nn\\
&=&(\Delta_2)_{xx}+{4\ov J\pi^2} \lambda\,g_2^2\,{1\ov J^3}\int\limits_{-1/2}^{1/2}dz\,
{\sin^4\pi zJ_1\ov z^2}\nn\\
&=&(\Delta_2)_{xx}+{2\ov J}\lambda'\,g_2^2\,x+O({1\ov J})
\eea
where the extra factor of ${1\ov J}$ comes from the square of the coefficients $c_{ix}$.

The case $A=\alpha$ is treated similarly. We find:
\bea
{\hat\gamma}_{\alpha\alpha}&=&(\Delta_2)_{\alpha\alpha}+{4\lambda' g_2^2 \ov \pi^2 J^3} {1-x\ov x} 
\sum_{m_i=-J/2}^{J/2}{\sin^4\pi{m_i\ov J}J_1}{{{m_i\ov J}+{n_\alpha\ov J_1}}\ov 
\left[{m_i\ov J}-{n_\alpha\over J_1}\right]^3}\nn\\
&=&(\Delta_2)_{\alpha\alpha}+{2\ov J}\lambda' g_2^2 {(1-x)} +O({1\ov J})
\eea
As before, the extra factor of $1/J$ comes from $c_{i\alpha}^2$.

Due to the extra $1/J$ factors, it follows that all corrections of order $g_2^2$ to the 
anomalous dimensions of double-trace operators vanish in the large $J$ limit. 
On the string theory side this corresponds to a vanishing $g_s^2$ correction to the
mass of 2-string states. This is indeed true, since these corrections come at 1-loop level
from contact terms and thus are of order $g_s^4$.


\section{Discussions}

We have discussed in detail the realization of the order $g_s$ corrections and  
some of the $g^2_s$ corrections to the string theory Hamiltonian in the gauge theory 
dual. For the operators we were interested in, i.e. operators with two scalar 
impurities, we have successfully recovered the string theory  prediction. Of 
crucial importance in recovering its matrix elements was the identification of 
an operator basis which is orthonormal to order $g_s^2$. While the mere fact that
the string theory Hamiltonian can be identified in gauge theory is not surprising, it 
is certainly a surprising result that its matrix elements are computable in perturbation 
theory. This extends to the interacting string theory the observation that its gauge theory 
dual develops small effective coupling. 

The explicit computations presented in this paper hold to order 
$g_s=\lambda' g_2$ and match the string theory predictions to this order, in the limit
$\mu\rightarrow\infty$. While on the string theory side it seems complicated to find 
the corrections of order ${1\ov \mu}$, this seems relatively straightforward in gauge 
theory, since they correspond to higher loop corretions at fixed genus. In particular, 
following the diagramatic expansion of the gauge theory, it is not hard 
to see that the $\gamma_{i A}$ entries of the anomalous dimension matrix have the 
following generic form
\be
\gamma_{i A}=\sum_{m\ge 0}\sum_{n\ge 0} a_{mn}(g_2\lambda'){}^{2n+1} 
\lambda'{}^{m}
\ee
The identification of the string theory Hamiltonian in the gauge theory allows one to 
approach this computation, as the string theory prediction is not know. The basis of 
operators found in this paper is the correct one for finding $a_{0m}$ for all positive $m$.

The basis-independent way of stating our main result is that the Yang-Mills version 
of the string Hamiltonian is the difference between the generator of scale 
transfomations and the $U(1)_R$ generator corresponding to the propagation 
of the plane wave. With hindsight, this result is not all that surprising. After all,
this is a consequence of the AdS/CFT dictionary, with a slight twist introduced 
by the light-cone gauge: the generator of scale transformations is the gauge 
theory dual of AdS time translations while the $U(1)_R$ generator is the gauge theory
dual of motions on one of the great circles on $S^5$. Whether this identification 
survives interactions depends on whether any of the global symmetry generators 
are broken at the quantum level. On the string theory side, $AdS_5\times S^5$ 
was argued to be an exact solution of string theory.
Thus, we do not expect its symmetry algebra to be modified. Therefore, the AdS/CFT
correspondence implies that 
\be
P^-=\Delta-J
\ee
to all orders in the string coupling constant. Matching the matrix elements on both 
sides becomes therefore a matter of  correctly  identifying the string state
- gauge theory operator map. However, this identification cannot be done without 
string theory input. One needs to pick some operator and fix the state operator 
map such that the gauge theory matrix elements agree with the string theory computation
in the desider basis. This generically fixes the freedom left by the requirement that the
scalar product of operators be diagonal. Then, the matching of the matrix elements 
of any other operator between gauge and string theory can be considered as a check of 
the duality at interaction level.

\

\

\noindent
{\bf\large Acknowledgements:}
We would like to thank  Joe Polchinski for usefull discussions.
The work of D.G. and A.M. was supported in part by the
National Science Foundation under Grant No. PHY99-07949.
The work of A.M. was partly supported by the RFBR
Grant No. 00-02-116477
and in part by the Russian Grant for the support of
the scientific schools No. 00-15-96557.
The work of R.R. was supported in part by DOE under 
Grant No. 91ER40618(3N) and in part by the National Science
Foundation under Grant No. PHY00-98395(6T).

\appendix{}
\noindent{\bf\Large Tree level 3-point functions}

In this appendix we compute the tree-level 3-point functions of BNM operators.

$\bullet~~~~~~~~
\langle{\bar {\cal O}}^{[\Phi,\Psi]_J}_{m}{\cal O}^{J_1}_{\Phi}{\cal O}^{J_2}_{\Psi}\rangle$

Start with
\be
{\cal O}^{J_1}_{\Phi}{\cal O}^{J_2}_{\Psi}=
Tr[\Phi Z^{J_1}]Tr[\Psi Z^{J_2}]
\ee
and project onto single trace operators. The second operator can be inserted 
between any two fields of the first operator and furthermore $\Psi$ can sit at any position 
in the string of $J_2$ $Z$ fields. Thus we find:
\be
{1\ov N}
\sum_{l=0}^{J_1}\sum_{n=0}^{J_2}Tr[\Phi Z^{l}(Z^{n}\Psi Z^{J_2-n})Z^{J_1-l}]~~.
\ee
Computing the scalar product between this and a single-trace operator $\cO_m^{[\Phi,\Psi]_J}$
we find that it is given by the double sum:
\be
\sum_{l=0}^{J_1}\sum_{n=0}^{J_2}
e^{2\pi i (l+n)\varphi_m^J}~~.
\ee
Indeed, the $l^{\rm th}$ term in $\cO_m^{[\Phi,\Psi]_J}$ is weighted with $exp(2\pi l \varphi_m^J)$
and $l$ is the number of $Z$ fields between $\Phi$ and $\Psi$. This sum can be eqsily computed
and it gives:
\be
\sum_{l=0}^{J_1}\sum_{n=0}^{J_2}
e^{-i (l+n)\varphi_m^J}={(1-e^{-i(J_1+1)\varphi_m^J})
(1-e^{i(J_1-1)\varphi_m^J})\ov (1-e^{-i\varphi_m^J})^2}~~.
\ee
Thus, to leading order in a $J\rightarrow\infty$ expansion is:
\be
\langle{\bar {\cal O}}^{[\Phi,\Psi]_J}_{m}{\cal O}^{J_1}_{\Phi}{\cal O}^{J_2}_{\Psi}\rangle=
-{g_2\ov\sqrt{J}}\,{\sin^2\pi m x\ov \pi^2 m^2}+{\cal O}({1\ov J^2})~~~~~~~~g_2={J^2\ov N}~,
~~x={J_1\ov J}
\ee

$\bullet~~~~~~~~\langle{\bar{\cal O}}^{[\Phi,\Psi]_J}_{m}{\cal O}^{[\Phi,\Psi]_{J_1}}_{n}
{\cal O}^{J_2}_{vac}\rangle$

Start with
\be
{\cal O}^{[\Phi,\Psi]_{J_1}}_{n}{\cal O}^{J_2}_{vac}={1\ov\sqrt{J_1J_2}}
\sum_{l=0}^{J_1}e^{2\pi il\varphi_n^{J_1}}Tr[\Phi Z^l\Psi Z^{J_1-l}]Tr[Z^{J_2}]
\ee
and project onto single trace operators. The result is:
\bea
&&~~~{1\ov N\sqrt{J_1J_2}}
\sum_{l=0}^{J_1}e^{2\pi il\varphi_n^{J_1}}J_2\times \PP_l~~~~~~~~~~
{\rm with ~~}\PP_l{\rm~~given~~by}\nn\\
&&\PP_l=(l+1)\times Tr[\Phi Z^{J_2+l}
\Psi Z^{J_1-l}]+ (J_1-l+1)Tr[\Phi Z^l\Psi Z^{J-l}]
\eea
where the overall factor of $J_2$ comes from the $J_2$ ways of choosing a $Z$ in 
$Tr[Z^{J_2}]$ and the factors of $l$ and $(J_1-l)$ in the bracket come from the number
of ways of inserting $Z^{J_2}$ in the first trace.

Computing the scalar product between this and the operator ${\cal O}^{[\Phi,\Psi]_J}_{m}$
we find:
\bea
&&{1\ov N}\sqrt{J_2\ov (J_1+1)(J+1)}
\sum_{l=0}^{J_1}e^{2\pi il\varphi_n^{J_1}}\left[(l+1)e^{2\pi i (J_1-l) \varphi_m^J}+(J_1-l+1)
e^{-2\pi i l \varphi_m^J}\right]\nn\\
&=&{1\ov N}{\sqrt{J_2\ov (J_1+1)\,(J+1)}}\left[
{\sin^2(\pi m x)\ov\sin^2\pi a} +J_1+2 {\sin(\pi m x)\cos\pi (mx+a)\ov\sin\pi a}\right]
\eea
where $a={m\ov J}-{n\ov J_1}$ and $x={J_1\ov J}$. The first two terms in the bracket 
above arise from the $l$ and $J_1-l$ terms while the last term comes from the $1$ in each 
bracket. 
Keeping the leading term in the limit $J,\,J_1\rightarrow\infty$, the final result is:
\be
\langle{\bar{\cal O}}^{[\Phi,\Psi]_J}_{m}{\cal O}^{[\Phi,\Psi]_{J_1}}_{n}
{\cal O}^{J_2}_{vac}\rangle=
{g_2\ov\sqrt{J}} {x^{3/2}\sqrt{1-x}} 
{\sin^2(\pi m x)\ov \pi^2 (mx-n)^2} + \cO({1\ov J})~~~~~~~~~~g_2={J^2\ov N}~,~~x={J_1\ov J}
\ee


\begin{thebibliography}{9}

\bibitem{BMN} D.~Berenstein, J.~Maldacena, H.~Nastase
 JHEP 0204 (2002) 013, hep-th/0202021

\bibitem{GMR}D.J.~Gross, A.~Mikhailov, R.~Roiban, hep-th/0205066

\bibitem{SAZA} A.~Santambrogio, D.~Zanon, hep-th/0206079

\bibitem{METS}  R.R.~Metsaev, Nucl.Phys. B625 (2002) 70, hep-th/0112044

\bibitem{METY}R.R.~Metsaev, A.A.~Tseytlin, Phys.Rev. D65 (2002) 126004, 
hep-th/0202109

\bibitem{MITH} N.R.~Constable, D.Z.~Freedman, M.~Headrick, 
S.~Minwalla, L.~Motl, A.~Postnikov, W.~Skiba, JHEP 0207 (2002) 017,
hep-th/0205089

\bibitem{SPVO} M.~Spradlin, A.~Volovich, hep-th/0204146

\bibitem{SPVOIIv3} M.~Spradlin, A.~Volovich, hep-th/0206073v3

\bibitem{SEST} C.~Kristjansen, J.~Plefka, G.W.~Semenoff, 
M.~Staudacher hep-th/0205033

\bibitem{GU} U.~G\"ursoy, hep-th/0208041

\bibitem{CHU}C.-S. Chu, V.V.~Khoze, M.~Petrini, R.~Russo, 
A.~Tanzini, hep-th/0208148

\bibitem{SESTII} N.~Beisert, C.~Kristjansen, J.~Plefka, G.W.~Semenoff, 
M.~Staudacher, hep-th/0208178

\bibitem{STRINGS} S.~Minwalla, talk at Strings2002.

\bibitem{Huang} M.-X Huang, hep-th/0205311

\bibitem{CHU0} C.-S. Chu, V.V.~Khoze, A.~Tanzini, JHEP {\bf 06} (2002) 011, hep-th/0206005

\bibitem{CHUI} C.-S. Chu, V.V.~Khoze, A.~Tanzini, JHEP {\bf 06} (2002) 011, hep-th/0206167

\bibitem{PARY} A.~Parnachev, A.V.~Ryzhov, hep-th/0208010

\bibitem{KOREANS} Y.~Kiem, Y.~Kim, S.~Lee, J.~Park, hep-th/0205279

\bibitem{HVER} H.~Verlinde, hep-th/0206059

\bibitem{MODOP} M.~Bianchi, B.~Eden, G.~Rossi, Y.~Stanev, hep-th/0205321


\end{thebibliography}
\end{document}